\pdfoutput=1
\documentclass[iop]{emulateapj}
\slugcomment{}
 
\shorttitle{Hub-filament system in the HII region G18.88-0.49}
\shortauthors{L.~K. Dewangan et al.}
\begin{document}
\title{New insights in the HII region G18.88$-$0.49: hub-filament system and accreting filaments}
\author{L.~K. Dewangan\altaffilmark{1}, D.~K. Ojha\altaffilmark{2}, Saurabh Sharma\altaffilmark{3}, 
S. del Palacio\altaffilmark{4}, N.~K. Bhadari\altaffilmark{1,5}, and A. Das\altaffilmark{6}}
\email{lokeshd@prl.res.in}
\altaffiltext{1}{Physical Research Laboratory, Navrangpura, Ahmedabad - 380 009, India.}
\altaffiltext{2}{Department of Astronomy and Astrophysics, Tata Institute of Fundamental Research, Homi Bhabha Road, Mumbai - 400005, India.}
\altaffiltext{3}{Aryabhatta Research Institute of Observational Sciences (ARIES), Manora Peak, Nainital, 263002, India.}
\altaffiltext{4}{Instituto Argentino de Radioastronom\'ia (CCT La Plata, CONICET; CICPBA; UNLP), C.C.5, (1894) Villa Elisa, Buenos Aires, Argentina.}
\altaffiltext{5}{Indian Institute of Technology Gandhinagar Palaj, Gandhinagar, 382355, India.}
\altaffiltext{6}{University of Hyderabad, Hyderabad - 500046, India.}
\begin{abstract}
We present an analysis of multi-wavelength observations of an area of 0$\degr$.27 $\times$ 0$\degr$.27 
around the Galactic H\,{\sc ii} region G18.88$-$0.49, which is powered by an O-type star (age $\sim$10$^{5}$ years). 
The {\it Herschel} column density map reveals a shell-like feature of extension $\sim$12 pc $\times$ 7 pc and mass $\sim$2.9 $\times$10$^{4}$ M$_{\odot}$ around the H\,{\sc ii} region; its existence is further confirmed by the distribution of molecular ($^{12}$CO, $^{13}$CO, C$^{18}$O, and NH$_\mathrm{3}$) gas at [60, 70] km s$^{-1}$.  
Four subregions are studied toward this shell-like feature, and show a mass range of $\sim$0.8--10.5 $\times$10$^{3}$ M$_{\odot}$. These subregions associated with dense gas are dominated by non-thermal pressure and supersonic non-thermal motions. 
The shell-like feature is associated with the H\,{\sc ii} region, Class~I protostars, and a massive protostar candidate, illustrating the ongoing early phases of star formation (including massive stars). The massive protostar is found toward the position of the 6.7 GHz methanol maser, and is associated with outflow activity. Five parsec-scale filaments are identified in the column density and molecular maps, 
and appear to be radially directed to the dense parts of the shell-like feature. This configuration is referred to as a ``hub-filament" system. Significant velocity gradients (0.8--1.8 km s$^{-1}$ pc$^{-1}$) are observed along each filament, suggesting that the molecular gas flows towards the central hub along the filaments. 
Overall, our observational findings favor a global non-isotropic collapse scenario as discussed in Motte et al. (2018), which can explain the observed morphology and star formation in and around G18.88$-$0.49.
\end{abstract}
\section{Introduction}
\label{sec:intro}
Understanding the formation process of massive OB-type stars ($\gtrsim$ 8 M$_{\odot}$) is still far from complete \citep{zinnecker07,tan14,Motte+2018}. Dust and molecular filaments are often associated with dense massive star-forming clumps, H\,{\sc ii} regions excited by massive stars, and young stellar clusters 
\citep[e.g.,][]{myers09,andre10,schneider12,Tige+2017,Motte+2018,morales19,kumar20}. 
It is thought that massive stars and clusters of young stellar objects (YSOs) commonly form within parsec-scale massive clumps/clouds such as hub-filament systems \citep[e.g., Monoceros R2;][]{morales19} and ridges (e.g., DR\,21). 
Hub-filament systems are known as a junction of three or more filaments \citep[e.g.,][]{myers09}. 
In general, in hub-filament systems, filaments are traced with high aspect ratio (length/diameter) and lower column densities ($\sim$10$^{21}$ cm$^{-2}$)
compared to a hub system that is identified with low aspect ratio and high column 
density \citep[$\sim$10$^{22}$ cm$^{-2}$; e.g.,][]{myers09,schneider12}. 

In relation to hub-filament systems, \citet{Motte+2018} discussed an evolutionary scheme for the birth of massive stars \citep[i.e., a global non-isotropic collapse scenario; see also][]{Tige+2017} that takes into account the flavours of the global hierarchical collapse and clump-feed accretion scenarios \citep[see][]{Vazquez-Semadeni+2009,Vazquez-Semadeni+2017,Smith+2009}. 
Central to such study is the identification of a hub-filament system containing the highest density regions, 
where several filaments converge. 
Such investigation requires the knowledge of molecular gas motion in the hub-filament system.
In this paper, we aim to identify embedded filaments and to investigate their role in the formation of massive OB-type stars 
and YSOs in the site hosting an H\,{\sc ii} region G18.88$-$0.49 \citep{kerton13} and an Extended Green Object \citep[EGO;][]{cyganowski08}. 

Following this section, an overview of G18.88$-$0.49 is presented in Section~\ref{sec:intro2}. 
Section~\ref{sec:obser} gives the information of the various data sets used in this paper. 
In Section~\ref{sec:data}, we present the outcomes concerning the physical environment and point-like sources toward G18.88$-$0.49. 
In Section~\ref{sec:disc}, we discuss possible star formation processes in G18.88$-$0.49. 
Finally, the major findings are summarized in Section~\ref{sec:conc}.
\section{Overview of G18.88$-$0.49}
\label{sec:intro2}
\begin{figure*}
\epsscale{0.6}
\plotone{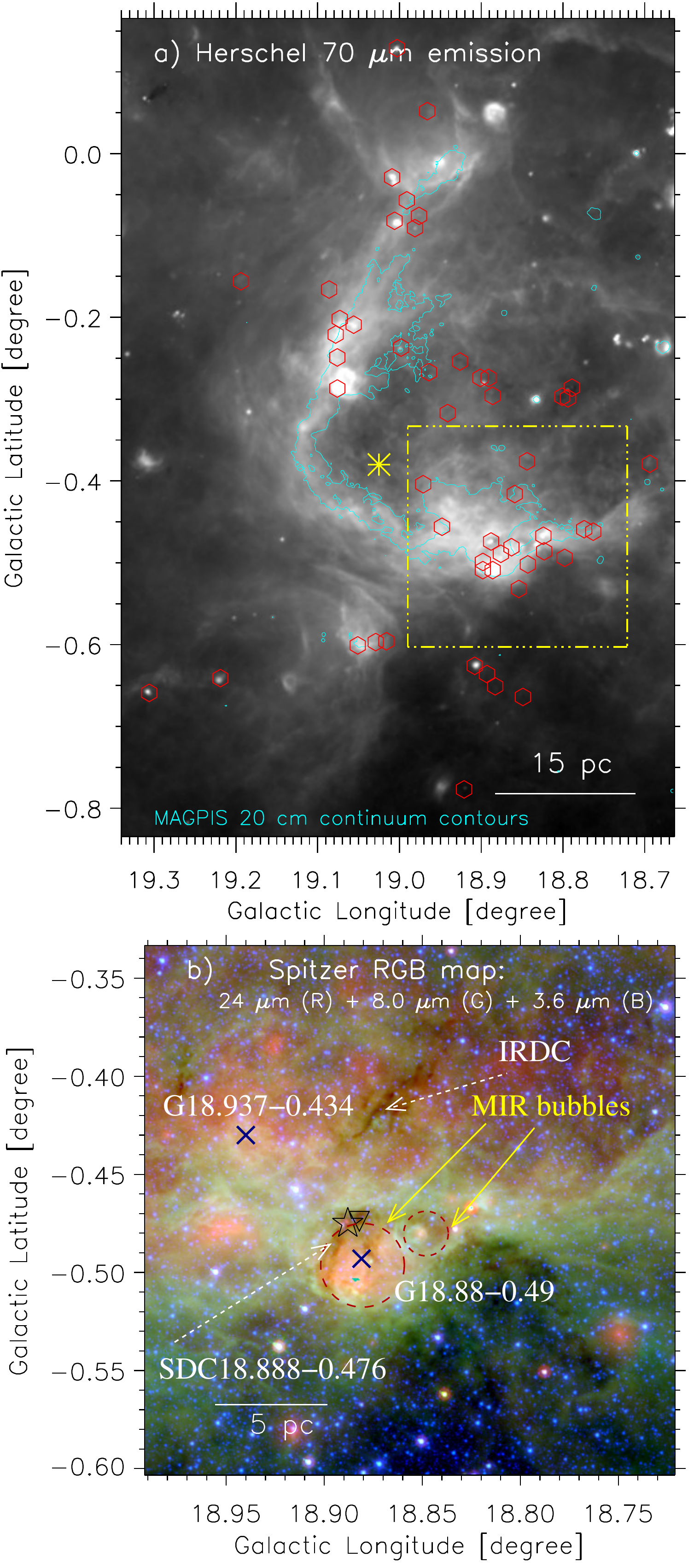}
\caption{a) The panel shows a large-scale view of the complex W39 (selected area $\sim$ 1$\degr$.2 $\times$ 1$\degr$.2 centered at 
{\it l} = 19$\degr$.176; {\it b} = $-$0$\degr$.333) using the {\it Herschel} 70 $\mu$m continuum image. 
The MAGPIS 20 cm continuum contour (in cyan) is also overplotted with a level of 3.0 mJy beam$^{-1}$. 
The ATLASGAL dust continuum clumps at 870 $\mu$m \citep[from][]{urquhart18} are also overlaid on the {\it Herschel} image (see hexagon symbols). 
An asterisk refers to the position of the exciting star(s) ({\it l} = 19$\degr$.025; {\it b} = $-$0$\degr$.38), 
which was previously reported by \citet{kerton13} \citep[see also][]{li19}. 
The dotted-dashed box (in yellow) encompasses the area shown in Figure~\ref{fig1}b, which is the target area of this paper. 
The scale bar referring to 15 pc (at a distance of 5.0 kpc) is shown.
b) Three color-composite map ({\it Spitzer} 24 $\mu$m (red), 8.0 $\mu$m (green), and 3.6 $\mu$m (blue) images in log scale) 
of an area ($\sim$0$\degr$.27 $\times$ 0$\degr$.27 centered at {\it l} = 18$\degr$.857; {\it b} = $-$0$\degr$.469) containing the H\,{\sc ii} regions G18.88$-$0.49 and G18.937$-$0.434. The position of the H\,{\sc ii} region G18.937$-$0.434 is obtained 
from \citet{anderson11}, while the position of the H\,{\sc ii} region G18.88$-$0.49 
is taken from \citet{lockman89} (see blue crosses). 
The scale bar referring to 5 pc (at a distance of 5.0 kpc) is shown. 
Dashed arrows show the locations of the IRDCs, while solid arrows and dashed circles highlight two MIR 
bubbles. 
The positions of a water maser and a 6.7 GHz methanol maser are marked by an upside down triangle and a star, respectively.} 
\label{fig1}
\end{figure*}

The selected H\,{\sc ii} region G18.88$-$0.49 is associated with the large, about 30 pc diameter, inner-Galaxy H\,{\sc ii} 
region W39 ({\it l} = 18$\degr$.7--19$\degr$.3) powered by a cluster of massive OB stars \citep{westerhout58,kerton13}. 
Figure~\ref{fig1}a displays a large-scale view of the complex W39 using the {\it Herschel} 70 $\mu$m continuum image 
overlaid with the 20 cm continuum emission contour from the Multi-Array Galactic Plane Imaging Survey \citep[MAGPIS;][]{helfand06}. 
This figure reveals a wide-open bubble and an extended H\,{\sc ii} region in the direction of W39 \citep[radial velocity (V$_\mathrm{lsr}$) range = (60, 70) km s$^{-1}$; distance $\sim$4.7 kpc; see][]{kerton13}. 
\citet{kerton13} found a local minimum in the radio emission that allowed them to propose the position of the exciting star(s) 
of the W39 H\,{\sc ii} region, which is indicated in Figure~\ref{fig1}a \citep[see also][]{li19}. 
Figure~\ref{fig1}a shows the positions of 52 APEX Telescope Large Area Survey of the Galaxy \citep[ATLASGAL;][]{schuller09} 
clumps at 870 $\mu$m from \citet{urquhart18}. 
These clumps are traced in a V$_\mathrm{lsr}$ range of $\sim$[59.5, 70] km s$^{-1}$ and a distance of $\sim$5.0 kpc \citep[see Table~2 in][]{urquhart18}. The H\,{\sc ii} region G18.88$-$0.49 is located toward the periphery of W39. 

The molecular cloud associated with the H\,{\sc ii} region G18.88$-$0.49 was reported by \citet{anderson09a}, and is 
referred to as C18.88$-$0.49 (V$_\mathrm{lsr}$ $\sim$65.5 km s$^{-1}$; distance $\sim$4.7 kpc). 
\citet{kerton13} reported that the H\,{\sc ii} region G18.88$-$0.49 is excited by a 
young O-type star (age $\sim$10$^{5}$ years). \citet{li19} studied the physical environment of W39 (including G18.88$-$0.49) using {\it Herschel} continuum images at 70--500 $\mu$m and the NH$_\mathrm{3}$ line data. 
To examine the near-infrared and mid-infrared (MIR) view, Figure~\ref{fig1}b shows a three color-composite image of G18.88$-$0.49 with {\it Spitzer} 
24 $\mu$m in red, 8.0 $\mu$m in green, and 3.6 $\mu$m in blue (see a dotted-dashed box in Figure~\ref{fig1}a). At least two MIR bubbles are seen in the {\it Spitzer} images, namely MWP-1G018879-004949 (radius = 1$'$.53) and MWP-1G018848-004761 (radius = 0$'$.88) (see arrows and dashed circles in Figure~\ref{fig1}b), which were previously reported in the Milky Way Project bubbles catalog \citep{simpson12}. 
In Figure~\ref{fig1}b, the positions of the radio recombination line (RRL) observations are marked \citep[taken from][]{lockman89,anderson11}. 
Apart from our selected target H\,{\sc ii} region, there is also another H\,{\sc ii} region, G18.937$-$0.434, labeled in Figure~\ref{fig1}b. 
The H\,{\sc ii} regions G18.88$-$0.49 and G18.937$-$0.434 are depicted in the ionized gas velocity of 65.5 and 68 km s$^{-1}$, respectively \citep[e.g.,][]{lockman89,anderson11}. 
Earlier, G18.937$-$0.434 was labeled as G018.937$-$0.434a (hydrogen RRL V$_\mathrm{lsr}$ $\sim$68 km s$^{-1}$) and G018.937$-$0.434b (hydrogen RRL V$_\mathrm{lsr}$ $\sim$36.2 km s$^{-1}$) in the source catalog of the Green Bank Telescope H\,{\sc ii} Region Discovery Survey \citep[GBT HRDS; e.g.,][]{anderson11}. 
\citet{wenger13} also studied the properties of helium and carbon RRL emission from the HRDS nebulae (including the H\,{\sc ii} region G018.937$-$0.434a). Concerning the source G018.937$-$0.434a, line widths of hydrogen and helium lines were reported to be 
23.83$\pm$0.05 and 16.31$\pm$0.85 km s$^{-1}$, respectively \citep{wenger13}, which are smaller than the ones observed in 
supernova remnants \citep[SNRs; i.e., $>$ 50 km s$^{-1}$;][]{liu19}. 
Based on the properties of RRL emission (i.e., the $^{4}$He$^{+}$/H$^{+}$ ionic abundance ratio and line widths), the G18.937$-$0.434 (or G018.937$-$0.434a) was classified as a Galactic H\,{\sc ii} region. In Figure~\ref{fig1}b, we find the absorption features against the 
Galactic background in the {\it Spitzer} images, indicating the presence of infrared dark clouds 
(IRDCs; see dashed arrows in Figure~\ref{fig1}b). 
A majority of the ATLASGAL clumps at a distance of 5.0 kpc are also seen in the direction 
of our selected area around G18.88$-$0.49 (see Figure~\ref{fig1}a). Hence, in this paper, we have adopted a distance of 5.0 kpc to the selected H\,{\sc ii} regions for further analyses.

An object studied earlier IRDC SDC18.888$–$0.476 (hereafter SDC18) from the catalogue of \citet{peretto09} is also highlighted in Figure~\ref{fig1}b. \citet{rigby18} presented the 1.2 and 2.0 mm continuum maps of SDC18, which were observed with the IRAM 30~m telescope. 
\citet{cyganowski08} reported an EGO at the centre of SDC18. In general, EGOs are always associated with the extended 
emission at 4.5 $\mu$m, and are potential candidates for massive protostellar outflows \citep[e.g.,][]{cyganowski08}. 
A water maser \citep[V$_\mathrm{lsr}$ $\sim$65.1 km s$^{-1}$;][]{walsh11} and a 6.7 GHz methanol maser \citep[V$_\mathrm{lsr}$ $\sim$56.4 km s$^{-1}$;][]{breen15,yang19} have also been observed toward SDC18, and the positions of these masers are 
marked in Figure~\ref{fig1}b. In the direction of SDC18, the position of the methanol maser coincides with that of the EGO. Additionally, noticeable YSOs were also reported around G18.88$-$0.49 \citep{kerton13}. Hence, star formation activity has been found toward the site around G18.88$-$0.49.

\citet{kerton13} reported an age difference between the W39 H\,{\sc ii} region (age $\sim$1.5 Myr) and the compact H\,{\sc ii} regions (i.e., G18.88$-$0.49 and G18.937$-$0.434; age $\sim$10$^{5}$ years). On the basis of the observed morphology of W39 and the age difference argument, they suggested a triggered star formation scenario in W39 \citep[see also][]{li19}. They proposed that the H\,{\sc ii} regions G18.88$-$0.49 and G18.937$-$0.434 were triggered by the expansion of the W39 H\,{\sc ii} region. This interpretation was also supported by \citet{li19}. 
However, it is an extremely difficult task to distinguish which stars are triggered and which ones formed spontaneously, and 
one should be cautious in interpreting observations in favour of triggered star formation \citep{dale15}. 

In this paper, we examine the region around G18.88$-$0.49 to search for a hub-filament morphology 
using a multi-wavelength approach. 
In this connection, we identify embedded filaments and explore their role in the star formation process in the H\,{\sc ii} region G18.88$-$0.49 
using the {\it Herschel} sub-millimeter continuum images and the molecular line data. 
We also study the nature of the H\,{\sc ii} regions G18.88$-$0.49 and G18.937$-$0.434 using the multi-frequency radio continuum data. 
\section{Data sets}
\label{sec:obser}
The size of the selected region is $\sim$0$\degr$.27 $\times$ 0$\degr$.27 centered at 
{\it l} = 18$\degr$.857; {\it b} = $-$0$\degr$.469, corresponding to a physical scale of $\sim$23.5 pc  $\times$ 23.5 pc 
at a distance of 5.0 kpc (Figure~\ref{fig1}a). We used the publicly available ``science-ready" observational data products from different surveys summarized in Table~\ref{tab1}. 

The $^{12}$CO(J =1$-$0), $^{13}$CO(J =1$-$0), and C$^{18}$O(J =1$-$0) line data were retrieved from 
the FOREST Unbiased Galactic plane Imaging survey with the Nobeyama 45-m telescope \citep[FUGIN;][]{umemoto17} survey, and are calibrated in main beam temperature ($T_\mathrm{mb}$) \citep{umemoto17}. 
The typical rms noise level\footnote[1]{https://nro-fugin.github.io/status/} is $\sim$1.5~K, $\sim$0.7~K, and $\sim$0.7~K for 
$^{12}$CO, $^{13}$CO, and C$^{18}$O lines, respectively. 
The multi-beam receiver, FOur-beam REceiver System on the 45-m Telescope \citep[FOREST;][]{minamidani16,nakajima19}, was utilized for the FUGIN survey. 
To improve sensitivities, we smoothed the FUGIN data cube with a Gaussian function, giving a spatial resolution of 35$''$.

We also use radio continuum maps at six different frequencies between 1 and 2 GHz (explicitly, at 1.06, 1.31, 1.44, 1.69, 1.82, and 1.95 GHz) from The HI/OH/Recombination line survey of the inner Milky Way (THOR) \citep{beuther16,bihr16,wang18}.

We also explore the {\it Herschel} temperature and column density maps (resolution $\sim$12$''$) produced for the {\it EU-funded ViaLactea project} \citep{molinari10b}. Both maps were generated using the Bayesian {\it PPMAP} procedure \citep{marsh15,marsh17} and downloaded from 
the publicly available site\footnote[2]{http://www.astro.cardiff.ac.uk/research/ViaLactea/}. 
 \begin{table*}
\scriptsize
\setlength{\tabcolsep}{0.025in}
\centering
\caption{List of different surveys utilized in this paper.}
\label{tab1}
\begin{tabular}{lcccr}
\hline 
  Survey  &  Wavelength/Frequency/line(s)       &  Resolution ($\arcsec$)        &  Reference \\   
\hline
\hline 
 Multi-Array Galactic Plane Imaging Survey (MAGPIS)                             & 20 cm                       & $\sim$6          & \citet{helfand06}\\
 The HI/OH/Recombination line survey of the inner Milky Way (THOR)                             & 1--2 GHz                       & $\sim$25          & \citet{beuther16}\\
 NRAO VLA Sky Survey (NVSS)                                   & 21 cm                       & $\sim$45          & \citet{condon98}\\
 FUGIN survey &  $^{12}$CO, $^{13}$CO, C$^{18}$O (J = 1--0) & $\sim$20        &\citet{umemoto17}\\
APEX Telescope Large Area Survey of the Galaxy (ATLASGAL)                 &870 $\mu$m                     & $\sim$19.2        &\citet{schuller09}\\
{\it Herschel} Infrared Galactic Plane Survey (Hi-GAL)                              &70--500 $\mu$m                     & $\sim$5.8--37         &\citet{molinari10}\\
{\it Spitzer} MIPS Inner Galactic Plane Survey (MIPSGAL)                                         &24 $\mu$m                     & $\sim$6         &\citet{carey05}\\ 
{\it Spitzer} Galactic Legacy Infrared Mid-Plane Survey Extraordinaire (GLIMPSE)       &3.6--8.0  $\mu$m                   & $\sim$2           &\citet{benjamin03}\\
\hline          
\end{tabular}
\end{table*}
\section{Results}
\label{sec:data}
\subsection{Physical environment of G18.88$-$0.49}
\label{sec:env}
\subsubsection{Multi-band view of G18.88$-$0.49}
\label{sec:clmtv}
\begin{figure*}
\epsscale{1}
\plotone{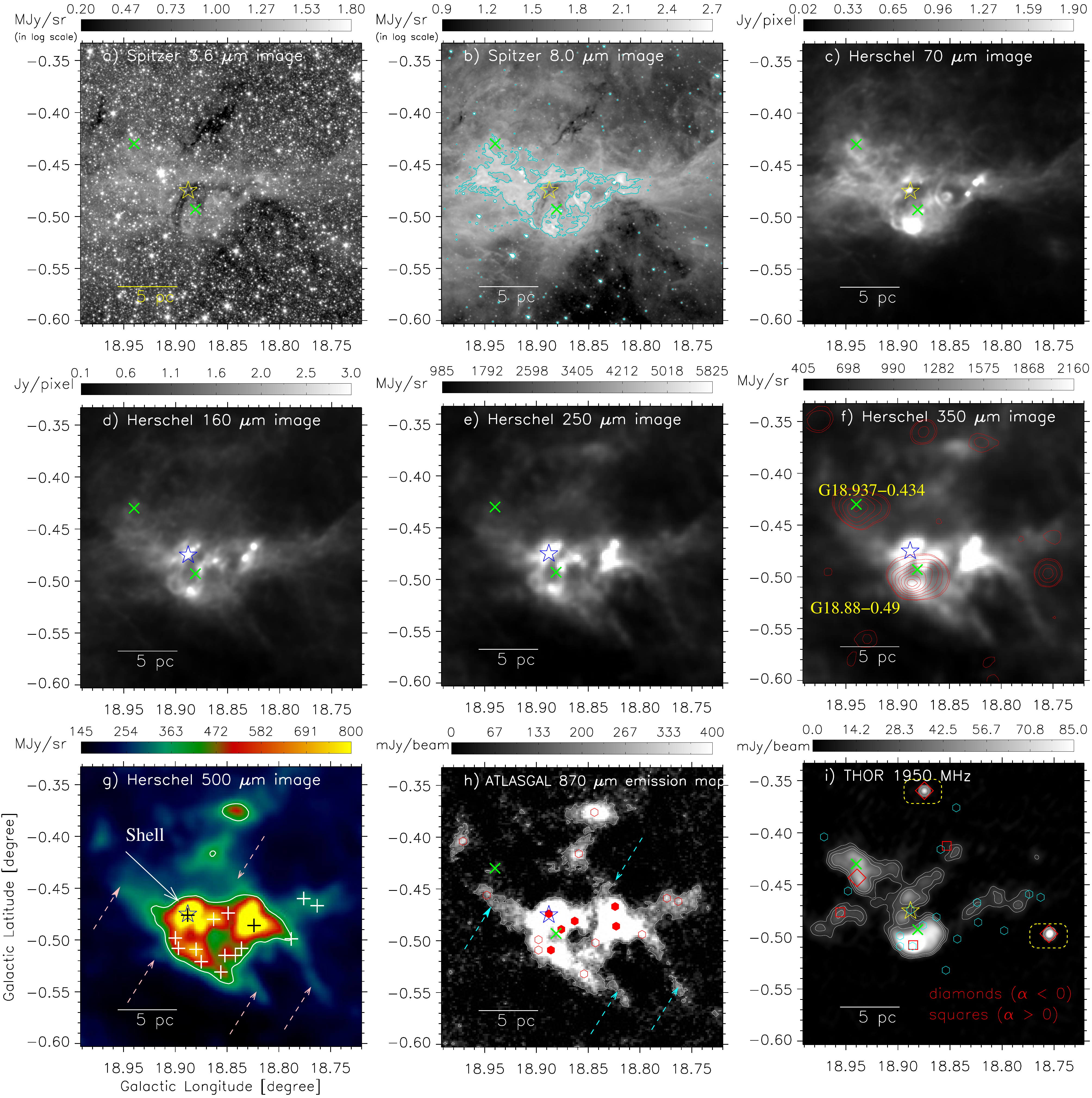}
\caption{Multi-wavelength view of the H\,{\sc ii} region G18.88$-$0.49. 
The images are displayed at different wavelengths, which are highlighted in the panels. 
Crosses indicate the position of the same H\,{\sc ii} regions as in Figure~\ref{fig1}b, 
whereas a star represents the position of a 6.7 GHz methanol maser. 
In panel ``b", the {\it Spitzer} 8.0 $\mu$m continuum contour is shown with a level of 230 MJy sr$^{-1}$. 
In panel ``f", the NVSS 1.4 GHz continuum contours (in red) are shown with the levels of 
5.45, 8 , 25, 50, 90, 130, 190, and 240 mJy beam$^{-1}$, where 1$\sigma$ $\sim$0.45 mJy beam$^{-1}$. 
In panel ``g", the {\it Herschel} 500 $\mu$m continuum contour (in white) indicating a shell-like morphology is shown with a level of 395 MJy sr$^{-1}$, and plus symbols show the positions of NH$_\mathrm{3}$ emission reported by \citet{li19}. 
In panel ``h", the ATLASGAL 870 $\mu$m continuum contour (in white) is shown with a level of 
125 mJy beam$^{-1}$. The ATLASGAL dust continuum clumps at 870 $\mu$m \citep[from][]{urquhart18} are also overlaid on the ATLASGAL image (see empty and filled hexagon symbols and also Figure~\ref{fig1}a). Physical parameters derived from the NH$_\mathrm{3}$ line data are available for six clumps \citep[filled hexagons; from][]{wienen12}. 
In panels ``g" and ``h", dashed arrows show potential filaments. 
In panel ``i", the THOR 1950 MHz continuum contours (in white) are shown with the levels of 12, 19, and 28 mJy beam$^{-1}$.
In panel ``i", open symbols (i.e., squares and diamonds) indicate the radio continuum sources from the THOR survey \citep{bihr16,wang18}; squares refer to the sources with spectral index $\alpha$ $>$ 0, and diamonds to sources with $\alpha$ $<$ 0. 
Two radio clumps appear to be extragalactic objects (see dashed boxes).} 
\label{fig2}
\end{figure*}

We examine the infrared, sub-millimeter (mm), and radio continuum images of G18.88$-$0.49 to 
investigate its physical environment (see Figure~\ref{fig2}). 
In Figures~\ref{fig2}a--\ref{fig2}i, we display images at 3.6 $\mu$m, 8.0 $\mu$m, 70 $\mu$m, 160 $\mu$m, 250 $\mu$m, 350 $\mu$m, 500 $\mu$m, 870 $\mu$m, and 1950 MHz, respectively. The positions of the 6.7 GHz methanol maser and the RRL observations are also marked 
in the multi-wavelength images. The IRDCs seen in the images at 3.6 and 8.0 $\mu$m appear as bright 
emission regions in the images at 70--870 $\mu$m. 
As highlighted earlier, SDC18 is associated with the 6.7 GHz methanol maser and the EGO. 
In the sub-mm images, we find a bright condensation/clump hosting the 6.7 GHz methanol maser/EGO. 
The {\it Herschel} and ATLASGAL images show the existence of potential filaments in our 
target site (see dashed arrows in Figures~\ref{fig2}g and~\ref{fig2}h). 
A shell-like feature is also seen in these images (see a contour and a solid arrow in Figure~\ref{fig2}g), and appears to be connected with potential filaments (see Sections~\ref{subsec:temp} and~\ref{sec:coem} for quantitative information). 
In the direction of the shell-like feature, the NH$_\mathrm{3}$ (1,1) emission has been observed \citep[see Figure~6 in][]{li19}, suggesting the presence of dense gas (see plus symbols in Figure~\ref{fig2}g). 
\subsubsection{Dust clumps}
\label{sec:dclmps}
The positions of 17 ATLASGAL dust clumps are also shown in Figure~\ref{fig2}h (see empty and filled hexagons). 
In Table~\ref{tab2}, we provide the positions and physical parameters of the clumps from \citet{urquhart18}. 
All these clumps are traced in a V$_\mathrm{lsr}$ range of $\sim$[61, 67.5] km s$^{-1}$. 
We also compute the average 
volume density of each clump as $n_{\mathrm H_\mathrm{2}}$ = 3M$_\mathrm{clump}$/(4$\pi$$R_\mathrm{clump}^{3}$$\mu_{\mathrm H_\mathrm{2}} m_{\mathrm H}$) (i.e., assuming that they are approximately spherical). 
Here, $R_\mathrm{clump}$ is the clump effective radius, $M_\mathrm{clump}$ is the clump mass, $m_{\mathrm H}$ is the mass of a hydrogen atom and the mean molecular weight $\mu_{\mathrm H_\mathrm{2}}$ is assumed to be 2.8. 
The values of $n_{\mathrm H_\mathrm{2}}$ for all the clumps are also tabulated in Table~\ref{tab2}, and 
vary between $\sim$2 $\times$ 10$^{2}$ and $\sim$1.27 $\times$ 10$^{4}$ cm$^{-3}$. 
We trace a dust temperature range of $\sim$13.7--25 K.
We find 8 clumps with $M_\mathrm{clump}$ $>$ 10$^{3}$ M$_{\odot}$ (see clump IDs with daggers in Table~\ref{tab2}). 

In Figure~\ref{fig2}h, we also highlight six clumps (see filled hexagons) for which their physical parameters were derived from the NH$_\mathrm{3}$ line data from \citet{wienen12}; such parameters are listed in Table~\ref{tab3}. 
Using these parameters, we also derive the sound speed ($a_\mathrm{s}$), thermal velocity dispersion ($\sigma_{T}$), non-thermal velocity dispersion ($\sigma_\mathrm{NT}$), Mach number ($M$ = $\sigma_\mathrm{NT}$/$a_\mathrm{s}$), and ratio of thermal to non-thermal gas pressure \citep[$R_\mathrm{p}$ = ${a^2_\mathrm{s}}/{\sigma^2_\mathrm{NT}}$; see][for more details]{lada03}. 
These parameters are also included in Table~\ref{tab3}. 
The value of the sound speed $a_\mathrm{s}$ (= $(k T_\mathrm{kin}/\mu m_\mathrm{H})^{1/2}$) is computed considering $\mu$ = 2.37 (approximately 70\% H and 28\% He by mass) and the kinematical temperature ($T_\mathrm{kin}$) as given in Table~\ref{tab3}. 
The non-thermal velocity dispersion is given by the following equation:
\begin{equation}
\sigma_\mathrm{NT} = \sqrt{\frac{\Delta V^2}{8\ln 2}-\frac{k T_\mathrm{kin}}{17 m_\mathrm{H}}} = \sqrt{\frac{\Delta V^2}{8\ln 2}-\sigma_\mathrm{T}^{2}} ,
\label{sigmanonthermal}
\end{equation}
where $\Delta V$ is the measured full width at half maximum (FWHM) linewidth of the observed NH$_\mathrm{3}$ spectra, 
$\sigma_\mathrm{T}$ (= $(k T_\mathrm{kin}/17 m_\mathrm{H})^{1/2}$) refers to the thermal broadening for NH$_\mathrm{3}$ \citep[e.g.][]{dunham11,dewangan16}. 
The values of $R_\mathrm{p}$ range from 0.04 to 0.12 with an average of 0.09 in six clumps, while the range of Mach numbers of these clumps is [2.9, 4.8] with an average of 3.5. These physical parameters suggest that non-thermal pressure and supersonic non-thermal motions (e.g., turbulence, outflows, shocks, and/or magnetic fields) are dominant in these clumps \citep[e.g.,][]{lada03}.
\subsubsection{Radio sources}
\label{sec:rclmps}
Figure~\ref{fig2}f displays the overlay of the NVSS 1.4 GHz continuum emission contours \citep[beam size $\sim$45$''$; noise level $\sim$0.45 mJy beam$^{-1}$;][]{condon98} on the {\it Herschel} 350 $\mu$m image. In the NVSS map, radio clumps are clearly seen toward the H\,{\sc ii} regions G18.88$-$0.49 and G18.937$-$0.434. 
However, no radio continuum peak is seen toward the position of the 6.7 GHz methanol maser. 
Using the {\it clumpfind} IDL program \citep{williams94}, we estimated the integrated fluxes at NVSS 1.4 GHz of the radio sources containing these H\,{\sc ii} regions. Using the integrated flux at NVSS 1.4 GHz, we computed the number of Lyman continuum photons ($N_\mathrm{UV}$) and 
the dynamical age ($t_\mathrm{dyn}$) of each H\,{\sc ii} region. Following the equations and analysis adopted in \citet{dewangan17a}, we compute the value of $\log{N_\mathrm{UV}}$ to be 
48.22 and 47.90 for the H\,{\sc ii} regions G18.88$-$0.49 and G18.937$-$0.434, respectively 
\citep[see][for equations]{matsakis76}. 
Based on these values, the H\,{\sc ii} regions G18.88$-$0.49 and G18.937$-$0.434 are excited by O9V--O9.5V and O9.5V--B0V  stars, respectively \citep[e.g.,][]{panagia73}. 
Considering the typical value of the initial particle number density of the ambient neutral gas 
($n_\mathrm{0}$ = 10$^{3}$ (10$^{4}$) cm$^{-3}$), the dynamical age of the H\,{\sc ii} regions G18.88$-$0.49 and G18.937$-$0.434 is estimated to be $\sim$0.4 (1.4) and 
$\sim$0.4 (1.3) Myr, respectively \citep[e.g.,][]{dyson80}. \citet{kerton13} also reported similar ages and spectral types of the powering stars of both the H\,{\sc ii} regions.
The electron temperature of 10$^{4}$~K, the isothermal sound velocity in the ionized gas (= 11 km s$^{-1}$; \citet{bisbas09}), 
and the distance of 5 kpc are adopted in this calculation.
The analysis assumes that the H\,{\sc ii} regions are uniform and spherically symmetric. 

The THOR survey provides six radio continuum maps at 1--2 GHz \citep[beam size $\sim$25$''$; noise level $\sim$0.3--1 mJy beam$^{-1}$;][]{bihr16} and spectral index of radio sources. 
The THOR maps at 1--2 GHz have a better resolution and sensitivity than the NVSS map at 1.4 GHz.
In Figure~\ref{fig2}i, we show the THOR 1950 MHz continuum map overlaid with 
the THOR 1950 MHz continuum contours. 
Figures~\ref{fig2}f and~\ref{fig2}i allow us to compare the observed radio emission of the H\,{\sc ii} regions G18.88$-$0.49 and G18.937$-$0.434. In the THOR 1950 MHz continuum map, we see several peaks that are not detected in the NVSS radio continuum map. 
We also do not find any radio continuum peak toward the position of the 6.7 GHz methanol maser in the continuum map at 1950 MHz. 
The positions of the ATLASGAL clumps are also highlighted in the THOR continuum map, allowing us to compare the positions of the dust clumps with the ionized emission.

In Figure~\ref{fig2}i, the positions of the radio continuum sources from the THOR survey are also marked. 
The spectral index is measured using the radio peak fluxes 
at 1.06, 1.31, 1.44, 1.69, 1.82, and 1.95 GHz \citep[e.g.,][]{bihr16}. 
The determination of the value of $\alpha$ helps us to infer whether the radio continuum emission is thermal or non-thermal \citep[e.g.,][]{rybicki79,longair92}. Thermally emitting sources show a positive or near zero spectral index. SNRs exhibit 
non-thermal emission with $\alpha$ $\approx$ $-$0.5, while extragalactic objects generally have a 
steeper $\alpha$ $\approx$ $-$1 \citep[e.g.,][]{bihr16}. Based on the values of $\alpha$, two THOR sources 
G18.875$-$0.360 ($\alpha$ $\sim-1.05$) and G18.755$-$0.497 ($\alpha$ $\sim-0.77$) are more likely to be extragalactic objects. 
The H\,{\sc ii} region G18.88$-$0.49 ($\alpha$ $\approx$ 0.09) shows thermal radio continuum emission, while 
the H\,{\sc ii} region G18.937$-$0.434 ($\alpha$ $\approx$ $-$0.4), powered by an O-type star, displays non-thermal radio continuum emission (see Section~\ref{sec:disc_nonthermal} for more details). 
Despite the value of $\alpha$ $\approx$ $-$0.4, the source G18.937$-$0.434 was reported as an H\,{\sc ii} 
region and not a SNR candidate \citep[see source G18.939$-$0.443 in][]{wang18}. 
As mentioned earlier, this source was also classified as a Galactic H\,{\sc ii} region \citep{anderson11,wenger13}. 
The observed spectral index of $\approx$ $-$0.4 is consistent with a combination of thermal and non-thermal emission. 
The estimation of the spectral type of the powering source of this H\,{\sc ii} region appears to be overestimated 
because the total radio continuum flux of the H\,{\sc ii} region has been treated as thermal 
emission (see also Section~\ref{sec:disc_nonthermal}).   
\subsubsection{{\it Herschel} temperature and column density maps}
\label{subsec:temp}
\begin{figure*}
\epsscale{1}
\plotone{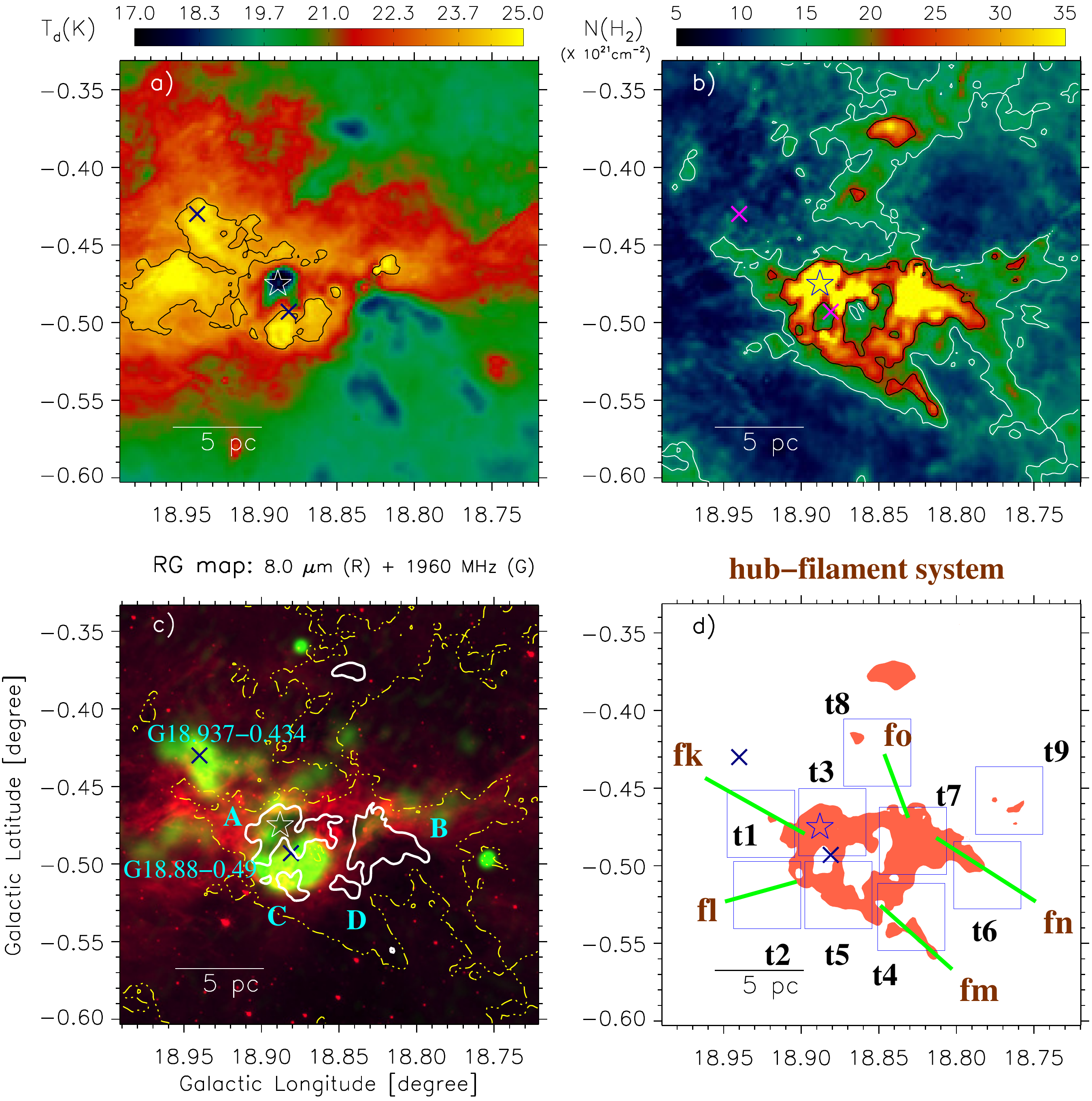}
\caption{a) {\it Herschel} temperature map. 
A solid contour (in black) is shown at $T_\mathrm{d}$ = 23.7 K in the figure. 
b) {\it Herschel} column density ($N(\mathrm H_2)$) map. 
Filamentary structures are traced in the column density map at a contour level of 14.1 $\times$ 10$^{21}$ cm$^{-2}$ (white contour). 
A solid contour (in black), at $N(\mathrm H_2)$ = 21 $\times$ 10$^{21}$ cm$^{-2}$, traces a shell-like feature in the figure.
c) Overlay of the $N(\mathrm H_2)$ contours on a 
two color-composite map (8.0 $\mu$m (red) and 1950 MHz (green) images). 
A dotted-dashed contour (in yellow) is drawn at $N(\mathrm H_2)$ = 14.1 $\times$ 10$^{21}$ cm$^{-2}$. 
A solid white contour at $N(\mathrm H_2)$ = 26.4 $\times$ 10$^{21}$ cm$^{-2}$ highlights the four subregions, A, B, C, and D. 
d) The panel shows a hub-filament system around the H\,{\sc ii} region G18.88$-$0.49. 
A filled contour displays the shell-like feature at $N(\mathrm H_2)$ = 21 $\times$ 10$^{21}$ cm$^{-2}$. 
A solid contour (in spring green) at $N(\mathrm H_2)$ = 26.4 $\times$ 10$^{21}$ cm$^{-2}$ shows the four subregions as labeled in Figure~\ref{fig3}c. 
Five filaments seen in the {\it Herschel} column density map are highlighted by solid lines (fk, fl, fm, fn, and fo), and are also labeled in the figure. 
Nine small regions (t1 to t9) are also indicated by boxes in the figure, where molecular spectra are extracted (see Figure~\ref{fig4}). 
In each panel, the scale bar, star, and crosses are the same as in Figure~\ref{fig1}b.}
\label{fig3}
\end{figure*}

Figure~\ref{fig3}a shows the {\it Herschel} temperature ($T_\mathrm{d}$) map of our selected target area around G18.88$-$0.49. We find the presence of warm dust emission ($T_\mathrm{d}$ = 23.5--25.5~K) 
toward both H\,{\sc ii} regions, G18.88$-$0.49 and G18.937$-$0.434. 
In Figure~\ref{fig3}b, we display the {\it Herschel} column density ($N(\mathrm H_2)$) map overlaid 
with the $N(\mathrm H_2)$ contours at [14.1, 21] $\times$ 10$^{21}$ cm$^{-2}$. 
Here, to trace different features in the {\it Herschel} column density map, we employ different $N(\mathrm H_2)$ contour levels through visual examination. The $N(\mathrm H_2)$ contour at 14.1 $\times$ 10$^{21}$ cm$^{-2}$ enables us to trace potential filaments in our selected target area. 
In the column density map, we identify the shell-like feature (extension $\sim$12 pc $\times$ 7 pc) at $N(\mathrm H_2)$ = 21 $\times$ 10$^{21}$ cm$^{-2}$ (see the black contour in Figure~\ref{fig3}b; and also Section~\ref{sec:clmtv}). The existence of the shell-like feature is also confirmed by the previously published NH$_\mathrm{3}$ (1,1) line data (see plus symbols in Figure~\ref{fig2}g), which is associated with the dense gas. With the help of the {\it Herschel} $N(\mathrm H_2)$ map, the total mass of the shell-like feature is determined to be $\sim$2.9 $\times$10$^{4}$ M$_{\odot}$ using the equation $M_{area} = \mu_{H_2} m_H A_{pix} \Sigma N(H_2)$, where $\mu_{H_2}$ is defined earlier (i.e., 2.8), $A_{pix}$ is the area subtended by one pixel (i.e., 6$''$/pixel), and $\Sigma N(\mathrm H_2)$ is the total column density \citep[see also][for more details of the analysis]{dewangan17a}. The {\it clumpfind} IDL program is employed to estimate the value of $\Sigma N(\mathrm H_2)$ \citep[e.g.,][]{dewangan17c,dewangan18b}. 

In Figure~\ref{fig3}c, we display a color-composite map with the {\it Spitzer} 8.0 $\mu$m and the THOR 1950 MHz images. One of the MIR bubbles (MWP-1G018879-004949 (radius = 1$'$.53)) surrounds the H\,{\sc ii} region G18.88$-$0.49. In other words, the MIR bubble is filled with the ionized gas. 
The color-composite map is also overlaid with the $N(\mathrm H_2)$ contours at [14.1, 26.4] $\times$ 10$^{21}$ cm$^{-2}$.  
At $N(\mathrm H_2)$ = 26.4 $\times$ 10$^{21}$ cm$^{-2}$, we identify at least four subregions around the H\,{\sc ii} region G18.88$-$0.49, which are labeled as A (mass $\sim$10.5 $\times$10$^{3}$ M$_{\odot}$), B (mass $\sim$8 $\times$10$^{3}$ M$_{\odot}$), C (mass $\sim$1.1 $\times$10$^{3}$ M$_{\odot}$), and D (mass $\sim$0.8 $\times$10$^{3}$ M$_{\odot}$). We followed the same steps as mentioned above to estimate the total mass of each sub-region. In Figure~\ref{fig3}d, we highlight five potential filaments (fk, fl, fm, fn, and fo) having typical lengths of five to eight parsecs, the shell-like feature, and four subregions (see also Section~\ref{sec:xcoem}). 
These proposed filaments, which have low column density, are associated with the shell-like feature having 
high column density. The association of the proposed filaments and the shell-like feature reveals an interesting configuration similar 
to a ``hub-filament" system \citep[e.g.,][]{myers09,schneider12,baug15,dewangan15,dewangan16,dewangan18a}. We find that the ends of the proposed filaments, which are approaching the hub, are 
associated with the dense parts of the shell-like feature. This finding is further explored using the analysis of the molecular line data 
in Section~\ref{sec:coem}. The H\,{\sc ii} region G18.88$-$0.49 excited by an O-type star is seen almost at the 
center of the shell-like feature, where no dust emission at $N(\mathrm H_2)$ = 21 $\times$ 10$^{21}$ cm$^{-2}$ is detected (see Figures~\ref{fig3}c and~\ref{fig3}d). 
These findings suggest that the energetics of a young massive O-type star (e.g., ionizing radiation, stellar winds, and radiation pressure) might have interacted with its immediate surroundings.
Additionally, some radio peaks are seen toward this shell-like feature (see subregions A, B and C). 
Furthermore, the H\,{\sc ii} region G18.937$-$0.434 does not appear to be part of the hub-filament system. 
\subsection{Morphology of molecular cloud and kinematics of molecular gas}
\label{sec:coem} 
\subsubsection{Moment and velocity channel maps}
\label{sec:xcoem} 
\begin{figure*}
\epsscale{1}
\plotone{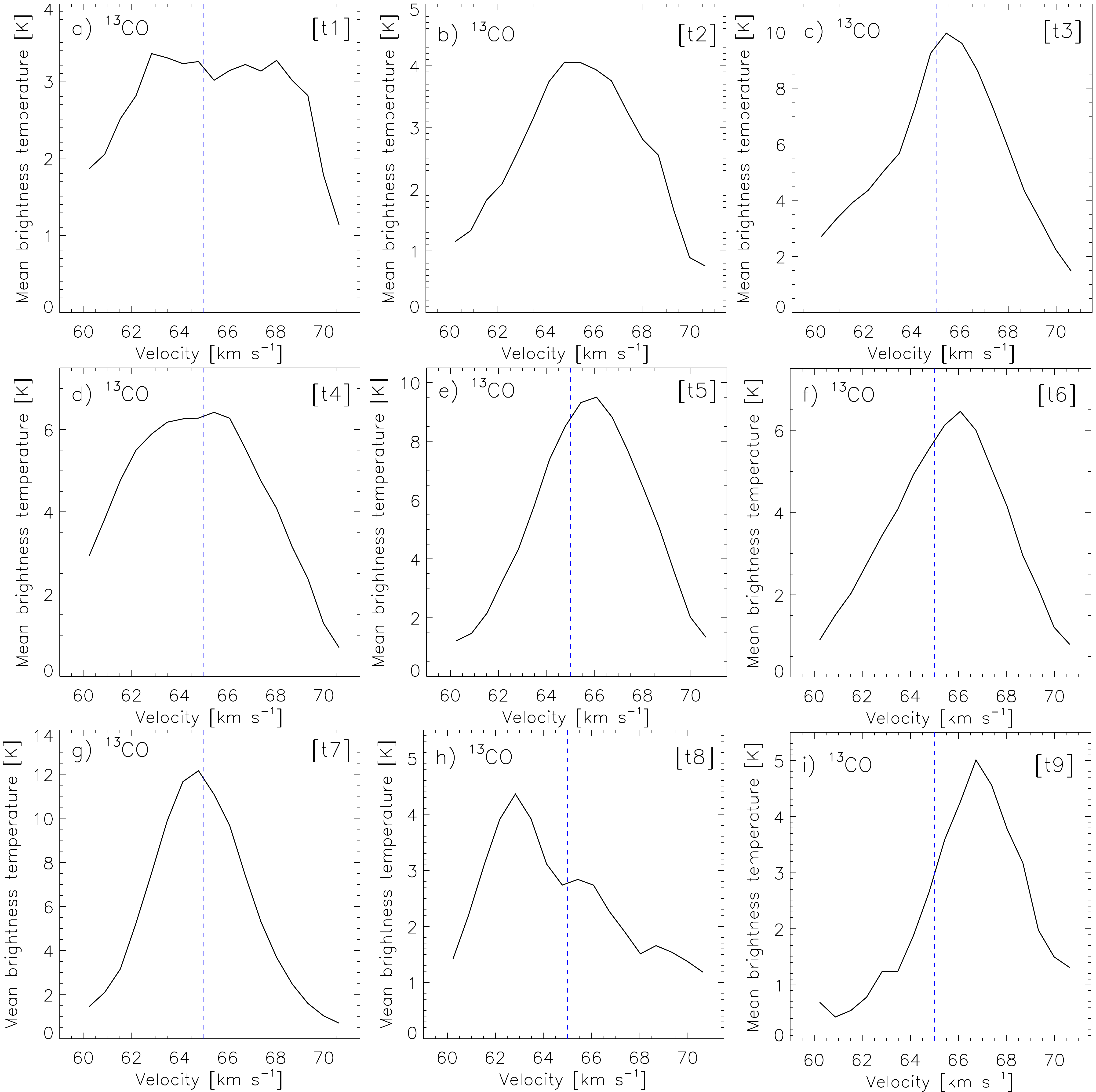}
\caption{$^{13}$CO profiles in the direction of nine small regions (i.e., t1 to t9; see corresponding boxes in Figure~\ref{fig3}d). In each panel, a dashed vertical line is drawn at V$_\mathrm{lsr}$ = 65 km s$^{-1}$.}
\label{fig4}
\end{figure*}

We extracted $^{13}$CO spectra toward nine small fields (i.e., t1--t9) that are highlighted in Figure~\ref{fig3}d. 
These are selected toward the regions around the shell-like feature. Figures~\ref{fig4}a--\ref{fig4}i present the averaged $^{13}$CO spectra toward nine small fields t1--t9. 
The spectra show that the molecular gas is well traced in a velocity range of [60, 70] km s$^{-1}$, confirming the existence of a single velocity component in the direction of our selected target site. 

\begin{figure*}
\epsscale{1}
\plotone{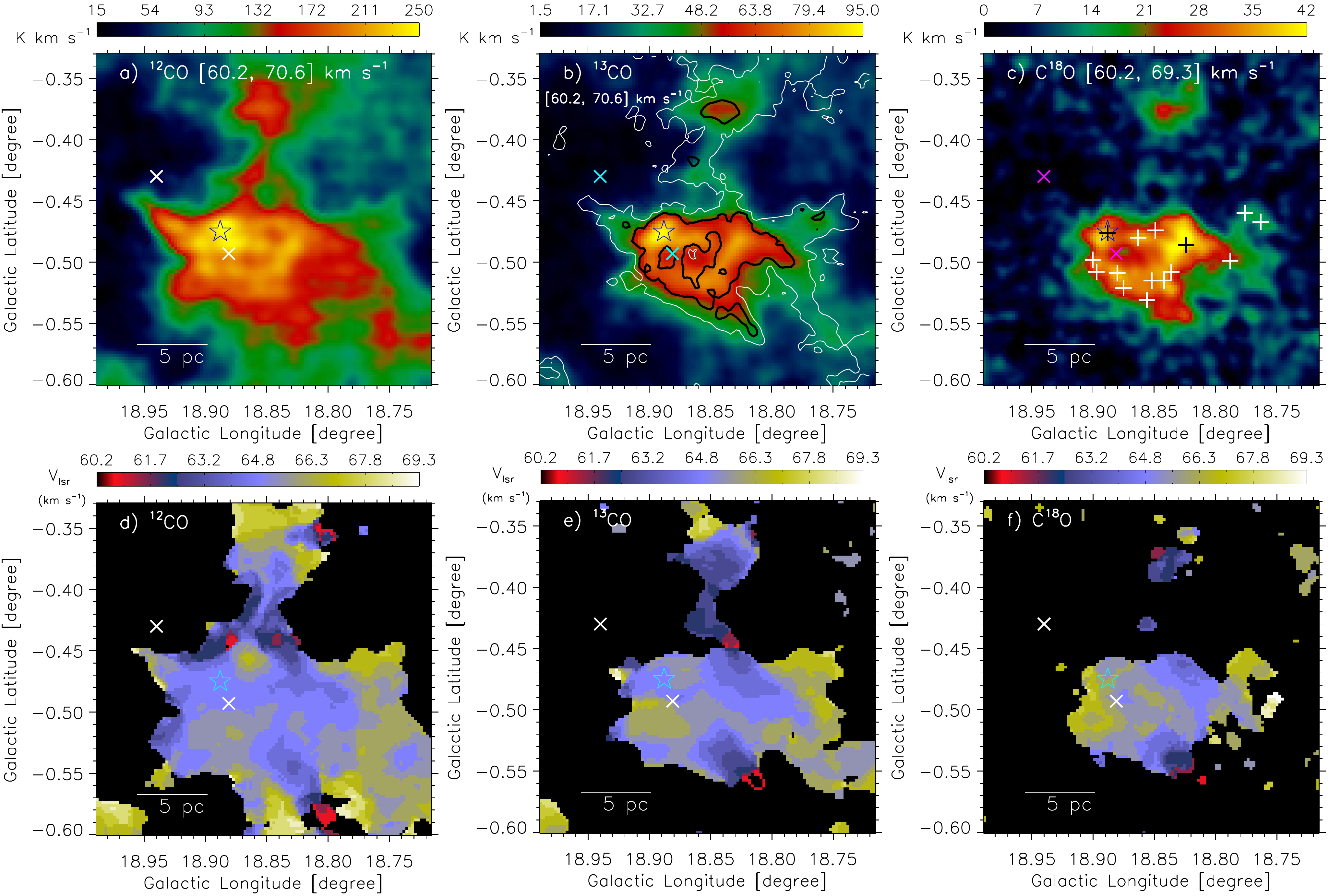}
\caption{FUGIN integrated intensity (or moment-0) maps of a) $^{12}$CO(J =1$-$0), b) $^{13}$CO(J =1$-$0), and c) C$^{18}$O(J =1$-$0) in the direction of the selected area around the H\,{\sc ii} region G18.88$-$0.49. FUGIN moment-1 maps of d) $^{12}$CO, e) $^{13}$CO, and f) C$^{18}$O.
In panels a), b), and c), the molecular emission is integrated over a velocity interval, which is given in each panel (in km s$^{-1}$). 
The scale bar, star, and crosses are the same as in Figure~\ref{fig1}b. 
The {\it Herschel} column density contours are also overlaid on the moment-0 map of $^{13}$CO (see Figure~\ref{fig5}b). 
A thin white contour is shown with the level of $N(\mathrm H_2)$ = 14.1 $\times$ 10$^{21}$ cm$^{-2}$, while a thick black contour is 
drawn at the level of $N(\mathrm H_2)$ = 21 $\times$ 10$^{21}$ cm$^{-2}$. In panel ``c", plus symbols highlight the positions of NH$_\mathrm{3}$ emission reported by \citet{li19}.}
\label{fig5}
\end{figure*}

To infer the morphology of the molecular cloud associated with the target site, the integrated $^{12}$CO, $^{13}$CO, 
and C$^{18}$O intensity maps (or moment-0 maps) are presented in Figures~\ref{fig5}a,~\ref{fig5}b, and~\ref{fig5}c, respectively. 
The molecular gas is integrated in a velocity range of [60.2, 70.6] km s$^{-1}$. 
In Figure~\ref{fig5}b, the column density contour at 14.1 $\times$ 10$^{21}$ cm$^{-2}$ is shown 
to highlight the proposed {\it Herschel} filaments. 
The black contour (at $N(\mathrm H_2)$ = 21 $\times$ 10$^{21}$ cm$^{-2}$) represents the shell-like feature in Figure~\ref{fig5}b. 
The $^{13}$CO emission follows quite well the morphology of the shell-like feature as investigated in the {\it Herschel} 
column density map. The moment-0 maps of $^{12}$CO and C$^{18}$O also trace the shell-like feature, 
which is also well depicted by the dense gas tracer NH$_\mathrm{3}$ (see plus symbols in Figure~\ref{fig5}c). 
The C$^{18}$O and NH$_\mathrm{3}$ emissions are not very strong toward the proposed filaments, 
indicating that these filaments are not very dense compared to the shell-like feature. 
However, low intensity $^{13}$CO emission is clearly seen toward all these filaments at $N(\mathrm H_2)$ = 14.1 $\times$ 10$^{21}$ cm$^{-2}$ (see a thin contour in Figure~\ref{fig5}b). No molecular emission is found toward the other H\,{\sc ii} region G18.937$-$0.434. 

The moment-1 maps (i.e., the intensity-weighted central velocity) of the $^{12}$CO, $^{13}$CO, and C$^{18}$O emission 
are shown in Figures~\ref{fig5}d,~\ref{fig5}e, and~\ref{fig5}f, respectively. 
Based on the comparison of the moment-1 maps to the integrated intensity maps, 
one can depict the noticeable velocity variations toward different parts of the shell-like feature. 
In Figures~\ref{fig5}e and~\ref{fig5}f, we see a noticeable sharp border between the two velocity fields toward the shell-like feature, revealing a velocity gradient.

\begin{figure*}
\epsscale{1}
\plotone{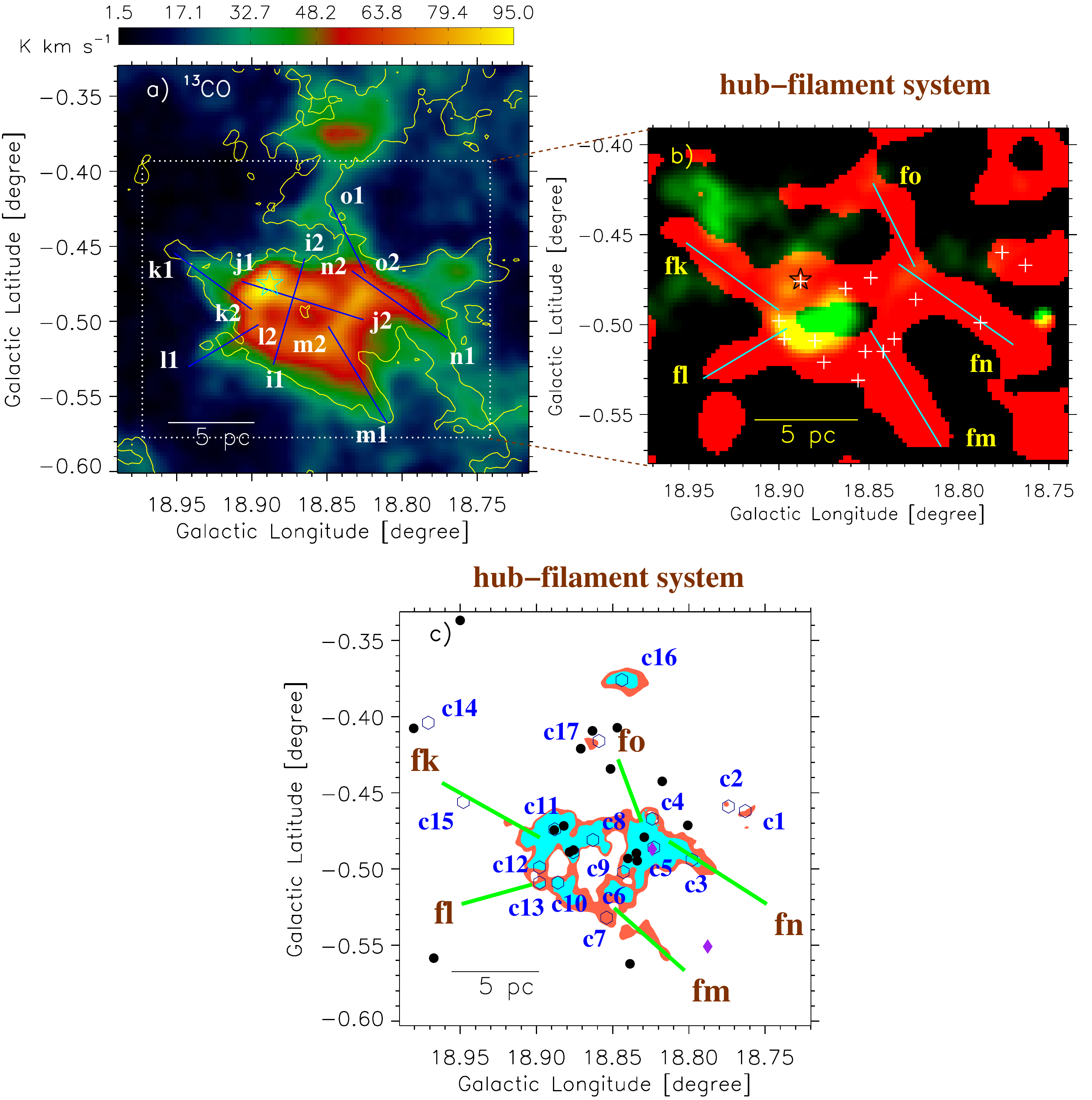}
\caption{a) FUGIN integrated intensity (or moment-0) map of $^{13}$CO (see also Figure~\ref{fig5}b).
Seven solid lines represent the axes (i.e., ``i1--i2", ``j1--j2", ``k1--k2", ``l1--l2", ``m1--m2", ``n1--n2", and ``o1--o2"), where position-velocity diagrams are extracted in 
Figures~\ref{fig10}a--~\ref{fig10}g. The dotted box (in white) encompasses the area shown in Figure~\ref{fig9}b. b) Overlay of the positions of NH$_\mathrm{3}$ emission \citep[see plus symbols; from][]{li19} on a two color-composite map (moment-0 map of $^{13}$CO (red) and 1950 MHz (green) images). 
The moment-0 map of $^{13}$CO is processed through an ``Edge-DoG" algorithm.  
c) The panel shows a hub-filament system around the H\,{\sc ii} region G18.88$-$0.49 as presented in Figure~\ref{fig3}d. 
The panel also displays the positions of Class~I YSOs (filled circles), flat-spectrum sources (filled diamonds), and the ATLASGAL clumps (empty hexagons). 
We identified Class~I YSOs and flat-spectrum sources using the {\it Spitzer} photometric data (see Figure~\ref{fig11} and also text for more details). A shell-like feature is shown by a filled tomato contour at $N(\mathrm H_2)$ =  21 $\times$ 10$^{21}$ cm$^{-2}$, and a filled cyan contour at $N(\mathrm H_2)$ = 26.4 $\times$ 10$^{21}$ cm$^{-2}$ highlights four subregions. In each panel, the scale bar is the same as in Figure~\ref{fig1}b. In panels ``a" and ``b", the position of a 6.7 GHz methanol maser is marked by a star.} 
\label{fig9}
\end{figure*}

In Figure~\ref{fig9}a, we highlight different axes (i.e., ``i1--i2", ``j1--j2", ``k1--k2", ``l1--l2", ``m1--m2", ``n1--n2", and ``o1--o2") in the integrated intensity map of $^{13}$CO. Along each of these axes, a position-velocity diagram is presented in Figure~\ref{fig10}. The axes ``i1--i2" and ''j1--j2" are selected in the direction of the major and minor axes of the shell-like feature, respectively. 
Figure~\ref{fig9}b presents a zoomed-in area around the shell-like feature using a color-composite map with the moment-0 map of $^{13}$CO and the THOR 1950 MHz images. Here, the moment-0 map of $^{13}$CO is processed through an edge detection algorithm \citep[i.e. Difference of Gaussian (DoG); see][]{gonzalez11,assirati14,dewangan17b}. It enables us to visually probe five filaments (i.e., fk, fl, fm, fn, and fo) as seen in the {\it Herschel} column density map (see Figure~\ref{fig3}d). The ionized gas is distributed almost at the center of the shell-like feature (extension $\sim$12 pc $\times$ 7 pc; aspect ratio $\sim$1.7), which is well traced with the NH$_\mathrm{3}$ emission (see plus symbols). The aspect ratios of the selected five molecular filaments vary between 2.3 to 4. Hence, this molecular map supports the existence of the hub-filament system around the H\,{\sc ii} region G18.88$-$0.49. Figure~\ref{fig9}c displays the hub-filament system overlaid with the positions of the ATLASGAL clumps, the shell-like feature, and the filaments. The positions of Class~I YSOs and flat-spectrum sources are also marked in Figure~\ref{fig9}c. The identification of these YSOs is discussed in Section~\ref{subsec:phot1}.

\begin{figure*} 
\epsscale{1}
\plotone{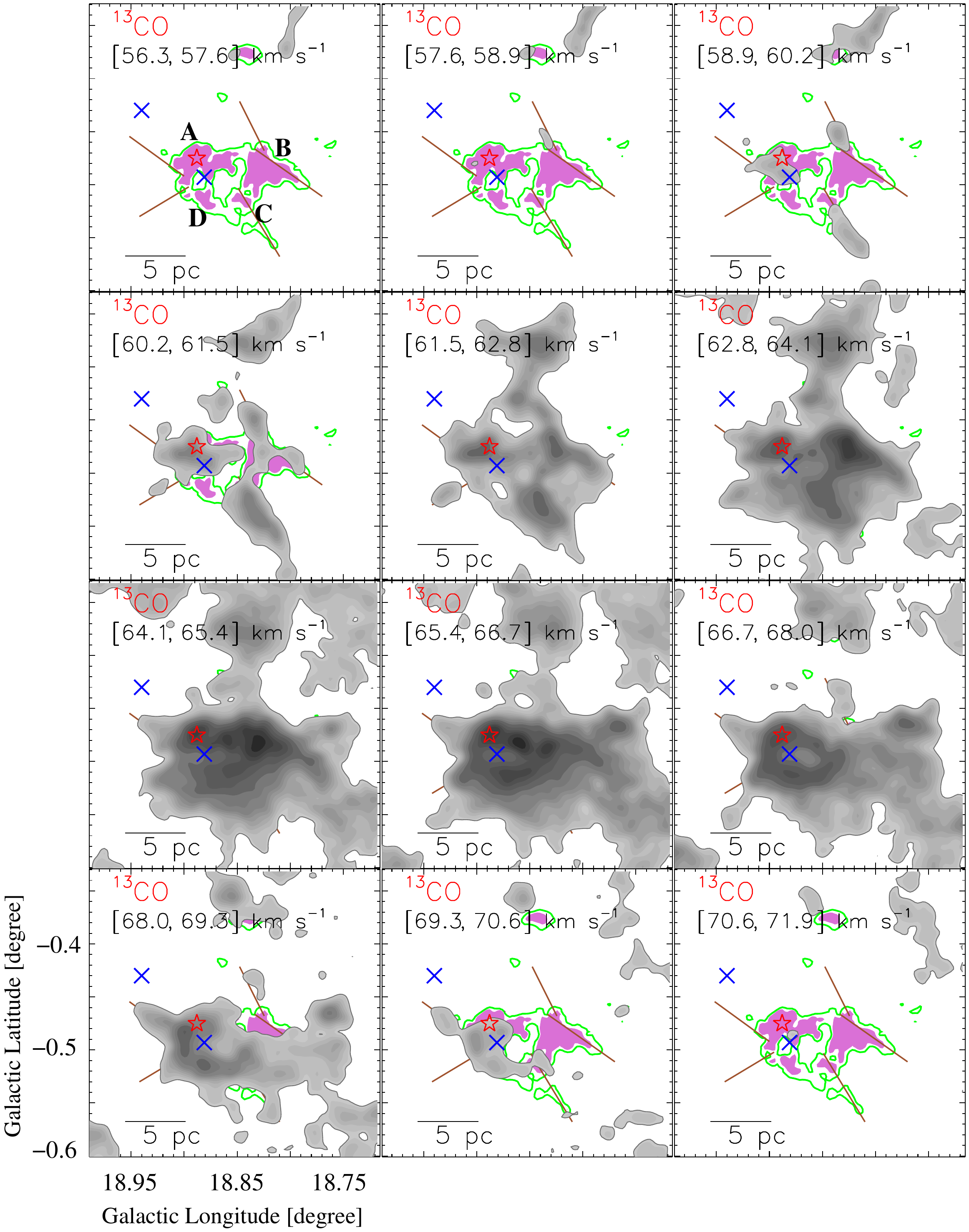}
\caption{Velocity channel maps of the $^{13}$CO emission.
The molecular emission is integrated over a velocity interval, which is given in each panel (in km s$^{-1}$). The contour levels are 4.5, 5, 6, 7, 8, 9, 10, 11, 13, 15, 18, 21, 23, 25, and 28 K km s$^{-1}$. In all panels, a shell-like feature is shown by a solid green contour at $N(\mathrm H_2)$ =  21 $\times$ 10$^{21}$ cm$^{-2}$, and a filled orchid contour at $N(\mathrm H_2)$ = 26.4 $\times$ 10$^{21}$ cm$^{-2}$ highlights the four subregions, A, B, C, and D. Five solid lines (in brown) represent the locations of the filaments in each panel (see also Figure~\ref{fig3}d). In each panel, the scale bar, star, and crosses are the same as in Figure~\ref{fig1}b.}
\label{fig6}
\end{figure*}

\begin{figure*} 
\epsscale{1}
\plotone{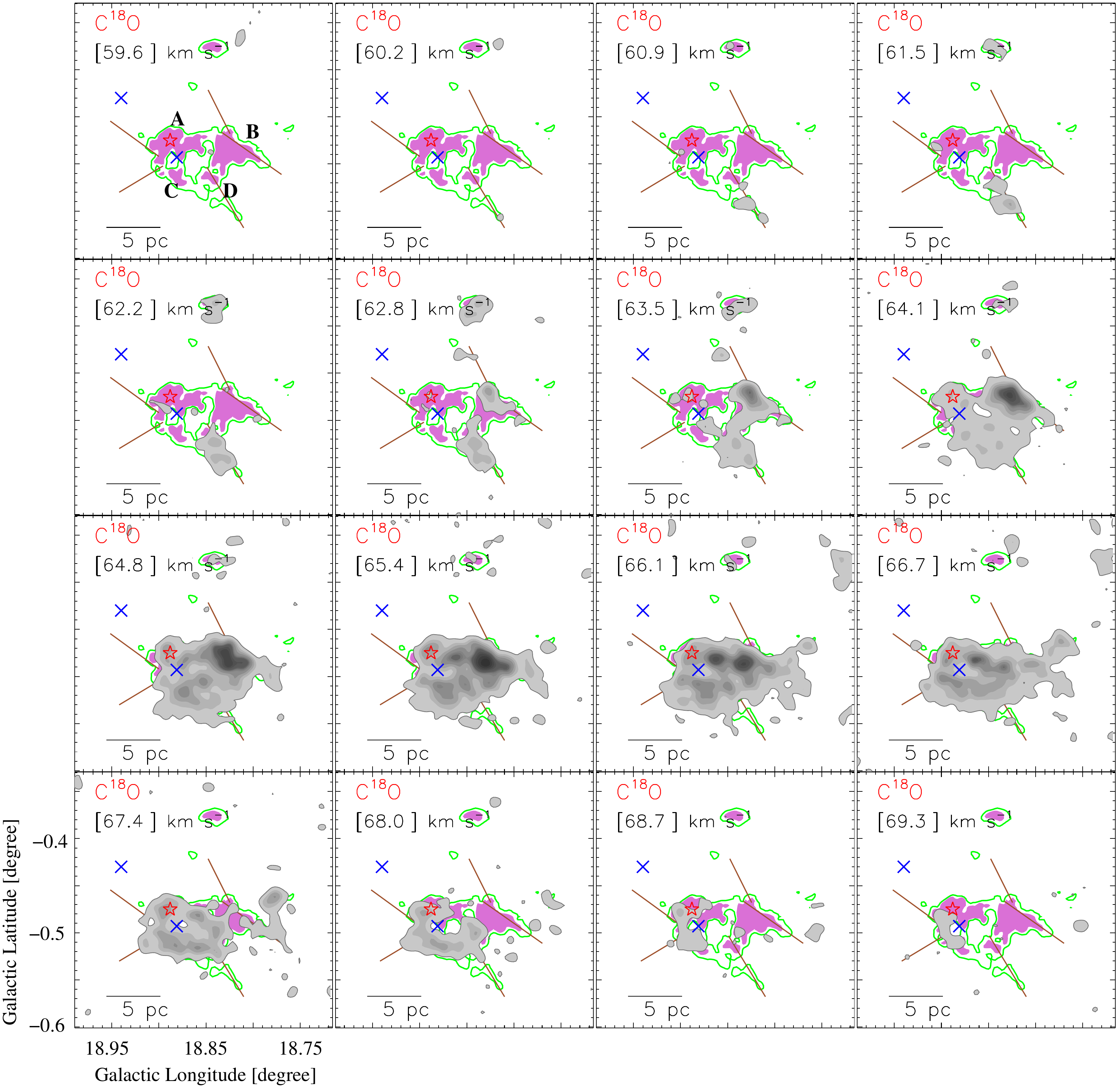}
\caption{Velocity channel maps of the C$^{18}$O emission. 
The velocity value (in km s$^{-1}$) is labeled in each panel. 
The contour levels are 1.2, 2, 2.5, 3, 3.5, 3.7, 3.9, 4.1, 4.6, 5, 5.5, 6, and 7.5 K. In all panels, a shell-like feature is shown by a solid green contour at $N(\mathrm H_2)$ =  21 $\times$ 10$^{21}$ cm$^{-2}$, and a filled orchid contour at $N(\mathrm H_2)$ = 26.4 $\times$ 10$^{21}$ cm$^{-2}$ highlights the four subregions, A, B, C, and D. Five solid lines (in brown) represent the locations of the filaments in each panel (see also Figure~\ref{fig3}d). 
In each panel, the scale bar, star, and crosses are the same as in Figure~\ref{fig1}b.}
\label{fig7}
\end{figure*}

Figure~\ref{fig6} displays the integrated $^{13}$CO velocity channel maps (starting from 56.3 km s$^{-1}$ at intervals of 1.3 km s$^{-1}$). 
In Figure~\ref{fig7}, we present the velocity channel maps of the C$^{18}$O emission (starting from 59.6 km s$^{-1}$). 
In Figures~\ref{fig6} and~\ref{fig7}, the shell-like feature and four {\it Herschel} subregions (``A--D") are also highlighted. 
Both channel maps trace the shell-like feature, and it appears that some parts of the shell-like feature appear clumpy (i.e., subregions). 
The $^{13}$CO and C$^{18}$O gas motion is also evident in the shell-like feature (including the {\it Herschel} subregions). 
The higher intensity of C$^{18}$O is found toward the sub-region ``B", while 
the sub-region ``A" is traced with the higher intensity of $^{13}$CO. 
In Figure~\ref{fig6}, the $^{13}$CO emission is also seen toward all the proposed filaments 
(see five solid lines and also the panel at [66.7, 68.0] km s$^{-1}$). 
\subsubsection{Position-velocity diagrams along filaments}
\label{sec:xxcoem} 
\begin{figure*}
\epsscale{0.92}
\plotone{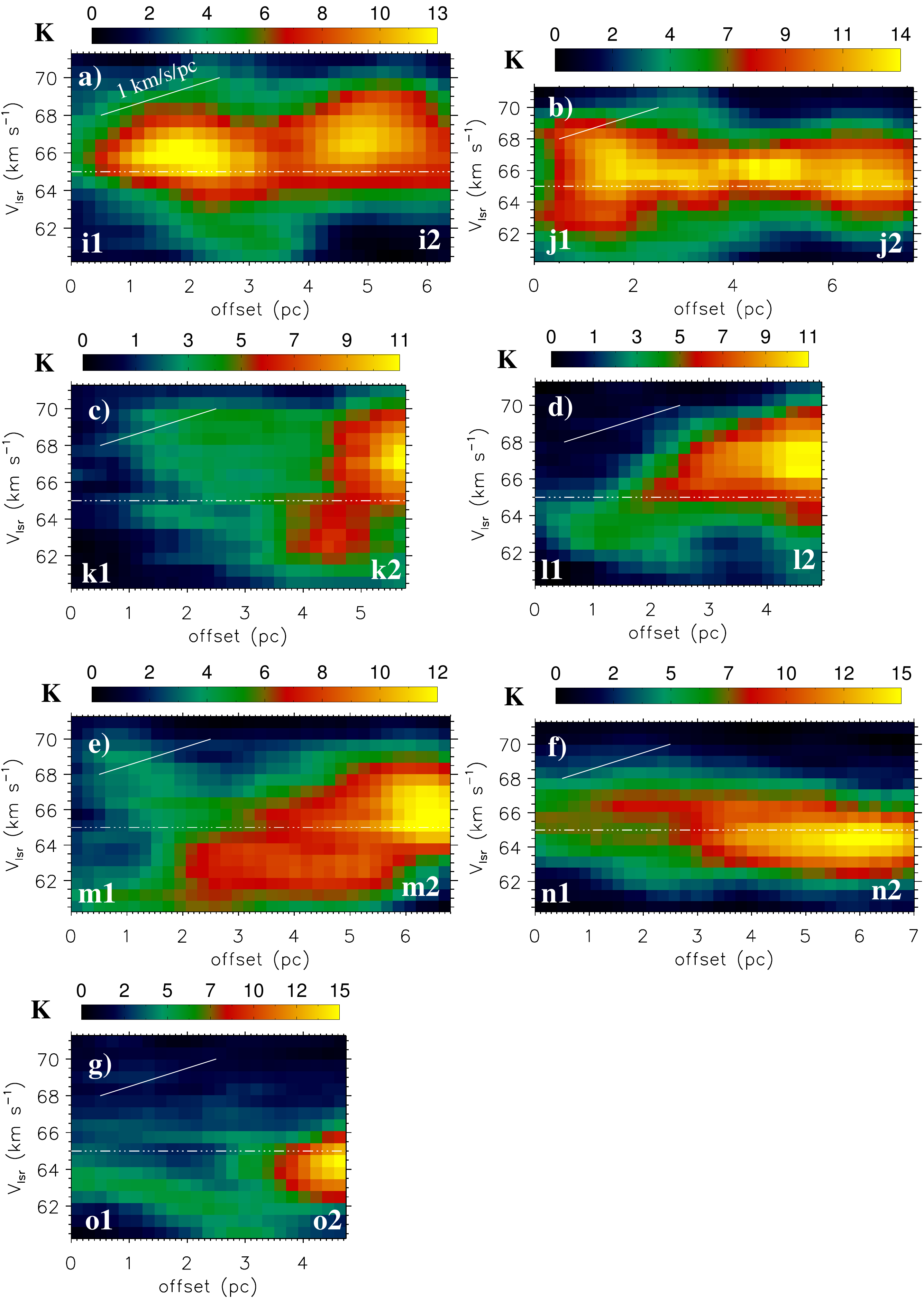}
\caption{Position-velocity diagrams along the axes a) ``i1--i2", b) ``j1--j2", c) ``k1--k2", d) ``l1--l2", e) ``m1--m2", f) ``n1--n2", and g) ``o1--o2" as marked in Figure~\ref{fig9}a. In each panel, a horizontal dotted-dashed line is shown at V$_\mathrm{lsr}$ = 65 km s$^{-1}$, and a reference bar at 1 km s$^{-1}$ pc$^{-1}$ is also drawn to facilitate the tracing of velocity gradients.}
\label{fig10}
\end{figure*}

In this section, we produce the position-velocity diagrams of $^{13}$CO to explore the velocity structure of the filaments. 
Figures~\ref{fig10}a,~\ref{fig10}b,~\ref{fig10}c,~\ref{fig10}d,~\ref{fig10}e,~\ref{fig10}f, and~\ref{fig10}g show the position-velocity diagrams of $^{13}$CO along the axes ``i1--i2", ``j1--j2", ``k1--k2", ``l1--l2", ``m1--m2", ``n1--n2", and ``o1--o2", respectively. 
In these axes, the positions k2, l2, m2, n2, and o2 are selected toward the central hub. 
In direction of the shell-like feature (see Figures~\ref{fig10}a and~\ref{fig10}b), a velocity spread is observed. 
Figures~\ref{fig10}c,~\ref{fig10}d,~\ref{fig10}e,~\ref{fig10}f, and~\ref{fig10}g show the existence of velocity 
gradients (about 0.8--1.8 km s$^{-1}$ pc$^{-1}$) along the filaments fk, fl, fm, fn, and fo, respectively. 
The end of each filament that reaches the hub is associated with the dense parts of the shell-like feature (see the positions k2, l2, m2, n2, and o2 in Figure~\ref{fig9}). The implication of the hub-filament configuration concerning the star formation process in G18.88$-$0.49 is presented in Section~\ref{sec:disc}.
\subsection{Star formation activity}
\label{subsec:phot1}
\begin{figure*}
\epsscale{1}
\plotone{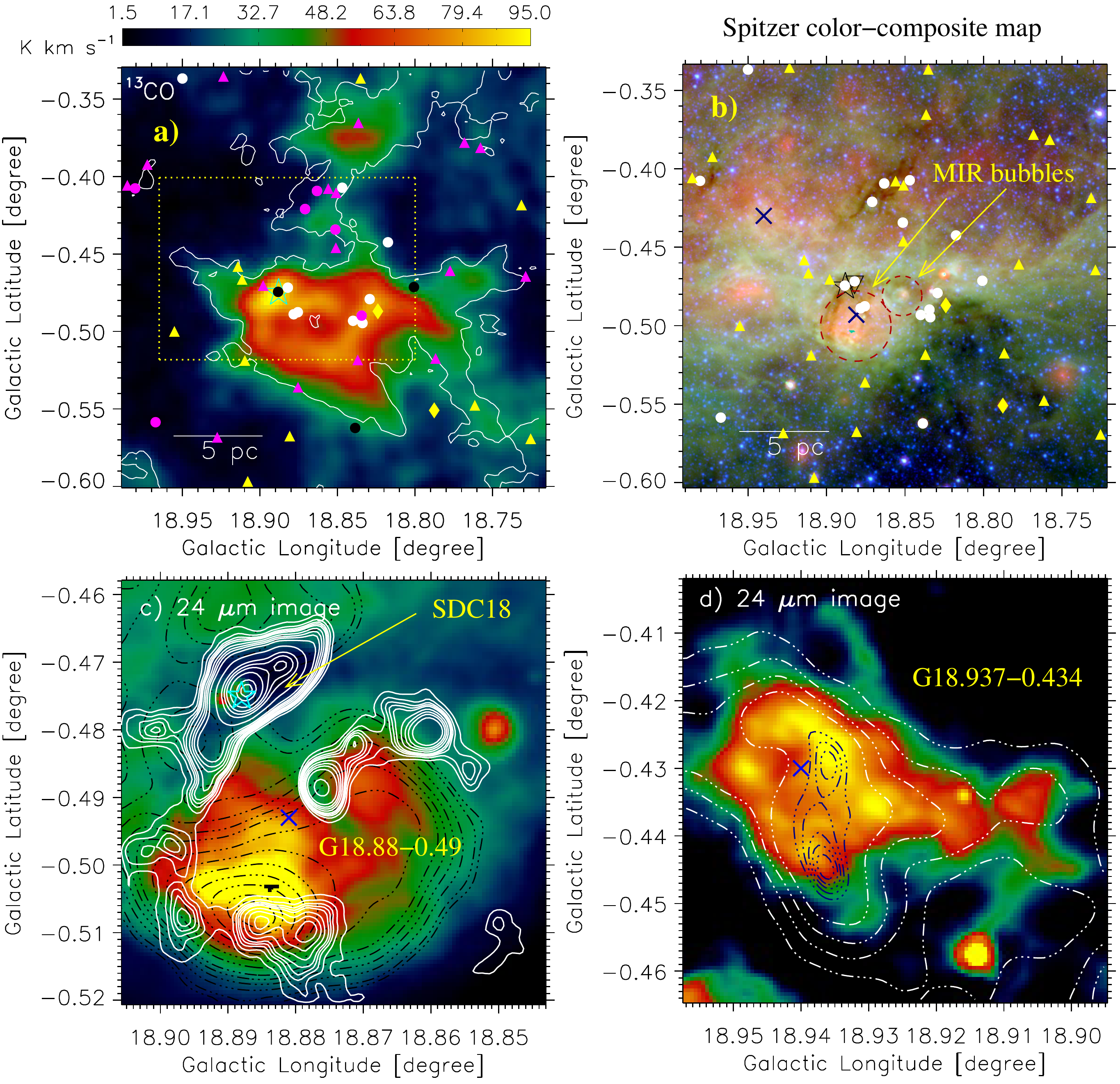}
\caption{a) Overlay of the positions of Class~I YSOs (filled circles), flat-spectrum sources (filled diamonds), and 
Class~II YSOs (filled triangles) on the integrated intensity map of $^{13}$CO. 
The symbols (in black and yellow) refer to the YSOs selected using the color-magnitude plot ([3.6] $-$ [24] vs [3.6]). 
The YSOs identified using the {\it Spitzer} color-color plot ([3.6]$-$[4.5] vs. [5.8]$-$[8.0]) are 
highlighted by symbols (in magenta), while the ones in white color are identified using the 
{\it Spitzer} color-color plot ([3.6]$-$[4.5] vs. [4.5]$-$[5.8]). 
A dotted yellow box highlights an area, where \citet{kerton13} reported the photometric magnitudes of their selected YSOs. 
The contour and the star symbol are the same as in Figure~\ref{fig9}a. 
b) Overlay of the positions of the selected YSOs (see filled circles, diamonds, triangles and also Figure~\ref{fig11}a) on the {\it Spitzer} color-composite map as shown in Figure~\ref{fig1}b.
c) A zoomed-in view of an area around the H\,{\sc ii} region G18.88$-$0.49 and SDC18 using the {\it Spitzer} 24 $\mu$m image. 
The map is also overlaid with the IRAM 1.2 mm continuum emission contours from \citet{rigby18} (see solid white contours) and the THOR 1950 MHz continuum contours (see dotted-dashed black contours). 
The contour levels of the IRAM 1.2 mm continuum emission are 28, 38, 45, 50, 55, 65, 75, 85, 100, 150, 220, 260, 300, 370, 500, and 570 mJy beam$^{-1}$, while the contour levels of the THOR continuum emission are (0.08, 0.1, 0.15, 0.2, 0.25, 0.3, 0.4, 0.5, 0.6, 0.7, 0.8, 0.9, 0.95) $\times$ 169 mJy beam$^{-1}$. 
d) A zoomed-in view of an area around the H\,{\sc ii} region G18.937$-$0.434 using the {\it Spitzer} 24 $\mu$m image. The map is also overlaid with the THOR 1950 MHz continuum contours (see dotted-dashed navy and white contours). 
The contour levels of the THOR continuum emission are 12, 22, 32, 42, 55, 58, 60, 61.5, and 63 mJy beam$^{-1}$. In panels, star and crosses are the same as in Figure~\ref{fig1}b.}
\label{fig11}
\end{figure*}

\citet{kerton13} identified YSOs using the {\it Spitzer} photometric data at 3.6--24 $\mu$m in W39, and provided the positions and photometric magnitudes of 36 YSOs (see Table~1 in their paper). 
We find only 19 out of 36 YSOs (6 Class~I; 7 Class I/II; 6 Class~II) that are restricted upto a small area of our selected target region (see a dotted yellow box in Figure~\ref{fig11}a). 
Hence, we also identified infrared-excess sources/YSOs in our entire selected area around G18.88$-$0.49 using the {Spitzer} 
photometric data at 3.6--24 $\mu$m. In this connection, three schemes are employed in this paper, which are 
color-magnitude plot ([3.6]$-$[24] vs [3.6]), 
color-color plot ([3.6]$-$[4.5] vs [5.8]$-$[8.0]), and color-color plot ([4.5]$-$[5.8] vs [3.6]$-$[4.5]). 
One can find more details of these schemes in \citet{dewangan17d,dewangan18a}. 
We obtained photometric magnitudes of point-like sources at {\it Spitzer} 3.6--8.0 $\mu$m bands from the GLIMPSE-I Spring' 07 highly reliable catalog. Furthermore, photometric magnitudes of sources at {\it Spitzer} 24 $\mu$m were also retrieved from the publicly 
available MIPSGAL catalog \citep[e.g.,][]{gutermuth15}.

Using the~color-magnitude~plot~([3.6]$-$[24] vs [3.6]; not shown here), different evolutionary stages of YSOs (i.e., Class~I, flat-spectrum, and Class~II) are identified against the possible contaminants \citep[i.e., galaxies and disk-less stars; see][]{guieu10,rebull11}. 
Note that this particular YSO identification and classification scheme was not used by \citet{kerton13}. 
With the application of this highlighted scheme, we find 15 YSOs (3 Class~I; 2 flat-spectrum; 10 Class~II) 
in our selected target area. In Figure~\ref{fig11}a, we overlay the positions of Class~I, flat-spectrum, and Class~II YSOs on the moment-0 map of $^{13}$CO, which are indicated by black filled circles, yellow filled diamonds, and yellow filled triangles, respectively.

Using the color-color plot ([3.6]$-$[4.5] vs [5.8]$-$[8.0]; not shown here), different evolutionary stages of 
YSOs (i.e., Class~I and Class~II) are selected against various contaminants (such as PAH-emitting galaxies and PAH-emission-contaminated apertures). Following the methods reported in \citet{gutermuth09} and \citet{lada06}, the selected YSOs and various contaminants are identified. With the help of this scheme, we find 22 YSOs (6 Class~I; 16 Class~II) in our selected target area. 
These selected YSOs are not common with the YSOs identified using the {\it Spitzer} color-magnitude plot ([3.6]$-$[24] vs [3.6]).
In Figure~\ref{fig11}a, the positions of the selected Class~I and Class~II YSOs are marked by magenta filled circles and triangles, respectively. 

Using the color-color plot ([4.5]$-$[5.8] vs [3.6]$-$[4.5]; not shown here), we identify Class~I YSOs with the infrared color conditions (i.e. [4.5]$-$[5.8] $\ge$ 0.7 mag and [3.6]$-$[4.5] $\ge$ 0.7 mag) given in \citet{hartmann05} and \citet{getman07} \citep[see also][]{dewangan20}. Using this scheme, we select 13 Class~I YSOs. 
These YSOs are not matched with the ones, which are identified in the above discussed two schemes.
One can notice that this highlighted scheme was not utilized by \citet{kerton13}. 
In Figure~\ref{fig11}a, the positions of the selected Class~I YSOs are indicated by white filled circles. 

Taken together, these three schemes yield a total of 50 YSOs (22 Class~I; 2 flat-spectrum; 26 Class~II) in our selected target field. 
In Figures~\ref{fig11}b, these YSOs are also overlaid on the {\it Spitzer} color-composite image. 
Hence, one can examine the spatial distribution of all these YSOs in Figures~\ref{fig11}a and~\ref{fig11}b. 
\citet{kerton13} identified 6 Class~I YSOs toward the area highlighted by a dotted yellow box in Figure~\ref{fig11}a, and 
5 out of these 6 Class~I YSOs are in agreement with the present work. 
However, the remaining one Class~I YSO is classified as a flat-spectrum source in this paper. 
No noticeable YSOs are seen toward the proposed filaments. We do not find YSOs toward the bubble MWP-1G018848-004761, while the Class~I YSOs, EGO, water maser, and 6.7 GHz methanol maser are seen toward the MIR bubble MWP-1G018879-004949 surrounding the H\,{\sc ii} region G18.88$-$0.49.  

There are 11 ATLASGAL clumps (c3--c13; see Table~\ref{tab2}), 8 Class~I YSOs, and 1 flat-spectrum source distributed toward the shell-like feature (see filled circles and filled diamonds in Figure~\ref{fig9}c). 
Furthermore, the 6.7 GHz methanol maser and the EGO are associated with the 
clump c11 ($L_{bol}$ $\sim$53.8 $\times$ 10$^{3}$ $L_\odot$; $M_\mathrm{clump}$ $\sim$4880 $M_\odot$; see Table~\ref{tab2}) and a Class~I YSO (GLIMPSE catalog ID: G018.8885$-$00.4746). 
In Figure~\ref{fig11}c, we display a zoomed-in view of an area around the H\,{\sc ii} region G18.88$-$0.49 and SDC18 
using the {\it Spitzer} 24 $\mu$m image. The image is also overlaid with the IRAM 1.2 mm continuum emission 
contours (resolution $\sim$ 13$''$) from \citet{rigby18} and the THOR 1950 MHz continuum contours. 
The Class~I YSO is also well detected in the {\it Spitzer} 24 $\mu$m image, and is seen toward the dust continuum peak at IRAM 1.2 mm. 
This source can be considered as an infrared counterpart (IRc) of the 6.7 GHz methanol maser emission, 
and is found toward the EGO \citep[e.g.,][]{cyganowski08}. 
Hence, this source is associated with the outflow activity. 
In general, the 6.7 GHz methanol maser ($<$0.1 Myr) is thought to be a reliable tracer of 
massive young stellar objects (MYSOs) \citep[e.g.][]{walsh98,minier01,urquhart13}. 
Hence, the IRc associated with the clump c11 may be a MYSO candidate \citep[stellar mass $\sim$7.7 M$_{\odot}$; see source ID \#G18-2 in Table~3 in][]{kerton13} going through an accretion phase. There is no radio peak seen toward the IRc (see Figure~\ref{fig11}c). Hence, it could be a genuine MYSO in a very early evolutionary stage, prior to an ultracompact (UC) H\,{\sc ii} phase. Note that the position of the H\,{\sc ii} region G18.88$-$0.49 is located away from that of the 6.7 GHz methanol 
maser (see Figure~\ref{fig11}c), suggesting the existence of different early evolutionary stages of massive star formation in the shell-like feature. 
It seems that other Class~I YSOs could be low-mass stars. The average age of Class~I YSOs has been reported to be $\sim$0.44 Myr \citep{evans09}.  

Together, the distribution of the selected Class~I YSOs, a MYSO candidate, and a young O-type star in the H\,{\sc ii} region G18.88$-$0.49 (age $\sim$10$^{5}$ years for $n_\mathrm{0}$ = 10$^{3}$ cm$^{-3}$) traces the early stage of star formation (including massive stars) 
toward the shell-like feature.
\section{Discussion}
\label{sec:disc}
\subsection{Star formation scenario in G18.88$-$0.49} 
\label{sec:disc11}
One of the important findings of the {\it Herschel} Space Observatory is the identification of numerous networks of filaments constituting hub-filament systems, which are gaining increasing attention as the most appropriate places for forming massive stars and star clusters \citep{Motte+2018,kumar20}. In the literature, several star-forming regions are reported to have a hub-filament system, namely Taurus \citep{myers09}, Ophiuchus \citep{myers09}, Rosette \citep{schneider12}, SDC335.579-0.292 \citep{peretto13}, IRDC G14.225$-$0.506 \citep{busquet13}, Sh 2-138 \citep{baug15}, W42 \citep{dewangan15}, IRAS 05480+2545 \citep{dewangan17b}, Sh 2-53 \citep{baug18}, and Monoceros~R2 \citep{morales19}. In the IRDC G14.225$-$0.506, \citet{busquet13} identified two hub-filament systems using the NH$_\mathrm{3}$ line data, and provided a table 
containing the aspect ratios of filaments and hubs to be 7.3--27.8  and 4.9--5, respectively (see Table~1 in their paper). Hence, in the hub-filament systems, hubs are more compact and dense compared to filaments.

It has also been proposed that hub-filament systems are the reliable sites for massive star formation, and 
the gas flows through the filaments are responsible for the central hub \citep[e.g.,][]{Tige+2017,Motte+2018,kumar20}. 
In the relation of ridges/hubs, \citet{Tige+2017} presented a scenario for the formation of massive 
stars \citep[see their Figure~8 and also][]{Motte+2018}. 
\citet{morales19} also discussed this scenario in support of their observational results in Monoceros~R2. 
Firstly, in this scheme, \citet{Tige+2017} and \citet{Motte+2018} highlighted a molecular cloud complex hosting a hub/ridge filament system with gas 
flowing via the filaments to the central hub, where massive dense cores/clumps (MDCs, in a 0.1~pc scale) develop.
During their starless phase ($\sim10^{4}$~yr), only low-mass prestellar cores are expected in MDCs. 
The MDCs become protostellar when containing a stellar embryo of low mass ($\sim3\times10^{5}$~yr). 
Then, the protostellar envelopes feed from the gravitationally-driven inflows, allowing the 
birth of massive protostars. Massive protostars become infrared-bright for stellar embryos with 
masses greater than 8 M$_{\odot}$. Finally, the main accretion phase halts when the stellar UV radiation ionizes the protostellar envelope and generates an H\,{\sc ii} region (in a time of few $10^{5}$--$10^{6}$~yr). 

The analysis of the {\it Herschel} column density map and the molecular line data has allowed us to uncover the physical environment of 
G18.88$-$0.49. 
These data sets show the existence of a massive shell-like feature (mass $\sim$2.9 $\times$10$^{4}$ M$_{\odot}$; extension $\sim$12 pc $\times$ 7 pc; aspect ratio $\sim$1.7) and 
five filaments (i.e., fk, fl, fm, fn, and fo; aspect ratio $\sim$2.3--4) having typical length of five to eight parsecs 
in the selected site around G18.88$-$0.49. It implies that filaments have high aspect ratio and lower column densities ($\sim$10$^{21}$ cm$^{-2}$), 
while the shell-like feature is associated with dense gas, and has low aspect ratio and high column density ($\sim$10$^{22}$ cm$^{-2}$). 
The physical association of the filaments and the shell-like feature reveals a hub-filament system around the H\,{\sc ii} region G18.88$-$0.49 
(see Sections~\ref{subsec:temp} and~\ref{sec:coem}).
These are the key and new results in G18.88$-$0.49. 

In this work we investigated four {\it Herschel} subregions toward the shell-like feature with masses of $\sim$0.8--10.5 $\times$10$^{3}$ M$_{\odot}$, 
which are also traced by the C$^{18}$O and NH$_\mathrm{3}$ emission (see Figure~\ref{fig5}c). 
In the direction of these subregions, we find dense clumps dominated by non-thermal pressure and supersonic non-thermal motions (see Sections~\ref{sec:dclmps} and~\ref{subsec:temp}). The shell-like feature is associated with the H\,{\sc ii} region G18.88$-$0.49, a MYSO candidate, and low-mass Class~I YSOs \citep[mean age $\sim$0.44 Myr;][]{evans09}. All these results reveal the ongoing star formation activity in the shell-like feature (see Section~\ref{subsec:phot1}). The center of the shell-like feature appears to be influenced by the energetics of the massive O-type star powering the H\,{\sc ii} region G18.88$-$0.49, which is surrounded by the MIR bubble MWP-1G018879-004949. 
However, this H\,{\sc ii} region (age $\sim$10$^{5}$ years) is not old enough to trigger a new generation of stars. 
Hence, this paper focuses to understand the role of observed features (i.e., filaments and shell-like feature) in physical processes involved in star formation in G18.88$-$0.49. 

We find five parsec-scale filaments that appear to be radially directed to the dense parts/subregions of the shell-like 
feature (see Figure~\ref{fig9}c). The FUGIN $^{13}$CO line data have been examined to study the velocity structure of the filaments (see Section~\ref{sec:coem}). Based on the position-velocity analysis of $^{13}$CO, we find noticeable velocity gradients 
(0.8--1.8 km s$^{-1}$ pc$^{-1}$) along five filaments (see Figure~\ref{fig10}). 
These velocity gradients are similar to those reported in NGC~1333 \citep[$\sim$0.5 -- 2.5 km s$^{-1}$ pc$^{-1}$;][]{hacar17}, Serpens South SF region \citep[1.4 km s$^{-1}$ pc$^{-1}$;][]{kirk13}, and 
Monoceros~R2 \citep[$\sim$0.0 -- 0.8 km s$^{-1}$ pc$^{-1}$;][]{morales19}. 
The observation of velocity gradients along molecular filaments is considered as a signpost of gas accreting along these filaments, which also feed star-forming cores and proto-clusters in the central hub \citep[e.g.,][]{kirk13,nakamura14,olmi16,morales19,chen20}. 
It has also been reported that the molecular gas might be accelerating when entering the hub \citep[e.g.,][]{morales19}. 
In the direction of the shell-like feature, our results show that the H\,{\sc ii} region G18.88$-$0.49 has been formed, while the IRc of the 6.7 GHz methanol maser (i.e., MYSO candidate) is being formed (see Figure~\ref{fig11}c). Hence, it is possible that the five filaments fk, fl, fm, fn, and fo had feeded and/or are feeding the central hub, 
where high $N(\mathrm H_2)$ materials ($\ge$ 21 $\times$ 10$^{21}$ cm$^{-2}$) and dense parts/subregions are observed.  
Our selected target system does not appear a classical ``hub-filament" system observed in 
Monoceros~R2 \citep{morales19}. Based on our observed findings, we can put our proposed hub-filament system in the phases of the 
infrared-bright massive protostar and the H\,{\sc ii} region as discussed in the scheme of \citet{Tige+2017} and \citet{Motte+2018}. 

Considering our observed outcomes, we find the applicability of a global non-isotropic 
collapse scenario in the H\,{\sc ii} region G18.88$-$0.49, which includes the flavours of the mentioned global hierarchical collapse and clump-feed accretion scenarios. Hence, the global non-isotropic collapse scenario can explain the observed morphology and star formation in and around G18.88$-$0.49.
\subsection{Radio emission in G18.937$-$0.434} 
\label{sec:disc_nonthermal}
The THOR and the NVSS radio continuum maps clearly trace the prominent H\,{\sc ii} 
region G18.937$-$0.434 (see Figures~\ref{fig2}f,~\ref{fig2}i, and~\ref{fig3}c). 
The {\it Herschel} maps show the presence of material with $N(\mathrm H_2)$ = 11.5--13.5 $\times$ 10$^{21}$ cm$^{-2}$ 
and warm dust ($T_\mathrm{d}$ = 23--25 K) around this H\,{\sc ii} region (see Figures~\ref{fig3}a and~\ref{fig3}b). 
An arc-like feature is also seen around this H\,{\sc ii} region in the {\it Spitzer} 8.0 $\mu$m 
image (see Figures~\ref{fig2}b and~\ref{fig3}c). In Figure~\ref{fig11}d, we present a zoomed-in view of an area around 
the H\,{\sc ii} region G18.937$-$0.434 using the {\it Spitzer} 24 $\mu$m image overlaid with the THOR 1950 MHz continuum contours. 
We find a noticeable MIR emission around the H\,{\sc ii} region G18.937$-$0.434, which coincides with the radio continuum emission.   
However, no noticeable molecular emission at [60, 70] km s$^{-1}$ is observed toward the H\,{\sc ii} region G18.937$-$0.434. 

Based on an examination of the catalog of THOR radio continuum sources in our selected target area, 
we find the H\,{\sc ii} region G18.937$-$0.434 with a negative spectral index ($\alpha$ $\approx$ $-$0.4; see Section~\ref{sec:rclmps}), which was 
produced using the radio peak fluxes at 1--2 GHz. 
We computed the integrated flux of G18.937$-$0.434 at each of the six THOR bands between 1 GHz and 2 GHz, and fitted the data with a single power-law component, 
leading to a spectral index $\alpha \approx -0.52$. One can consider about 10\% error in the values of flux and $\alpha$. The negative $\alpha$ value indicates non-thermal (synchrotron) radio emission from G18.937$-$0.434, thus revealing the 
presence of relativistic electrons. Note that we expect a composite spectrum including a thermal component as well, likely produced by the 
ionized gas, but the limited bandwidth of the radio observations does not allow us to properly disentangle these emission components. 
Nonetheless, the non-thermal emission has a negative spectral index and it should dominate at low frequencies, whereas the thermal emission 
is more prominent at higher frequencies. Hence, to interpret the detection of synchrotron emission we focus on the 
flux value at the lowest frequency accessible in the THOR survey, $S_{1.06~\mathrm{GHz}} = 1.135$~Jy, as the thermal contamination should be lowest in this band. Assuming that the thermal emission is optically thin (spectral index equal to $-$0.1) 
and that the non-thermal emission has a spectral index between $-$0.5 and $-$1, the thermal emission at 1.06 GHz is 
estimated to be $<$ 0.2 -- 0.6 Jy. If we consider the total thermal flux $S_{1.06~\mathrm{GHz}}$ = 0.2(0.6) Jy of 
the H\,{\sc ii} region G18.937$-$0.434 then it can be excited by a spectral type of B0.5V--B0V (B0V--O9.5V) star 
(see Section~\ref{sec:rclmps} for the analysis details). Hence, we suggest that 
the H\,{\sc ii} region G18.937$-$0.434 is powered by an early B-type main-sequence star. 

We adopt the same approach as \citet{dewangan20a} for modelling the non-thermal radio emission from an H\,{\sc ii} region. 
Taking as a reference a non-thermal spectral index of $\sim$ $-$0.8 as \citet{dewangan20a}, 
the non-thermal flux at 1.06 GHz is $\sim$0.7 Jy. We set the ratio between the energy density in relativistic protons and electrons to 100 and assume a minimum-energy condition for the non-thermal particles and the magnetic field. 
Our analysis yields an energy density of non-thermal particles of $U_\mathrm{NT} \sim 34$~eV~cm$^{-3}$ and a magnetic field intensity in the H\,{\sc ii} region of $\sim 43~\mu$G. 
The required value of $U_\mathrm{NT}$ greatly exceeds the Galactic cosmic ray energy density ($U_\mathrm{GCR} \approx 1$~eV~cm$^{-3}$). We conclude that the observed synchrotron flux is produced by relativistic particles locally accelerated in shocked regions of gas, consistent with recent non-thermal emission models \citep{padovani19}. 
However, we still need more observations in frequency ranges below 1 GHz and above 2 GHz to make more robust estimates of the magnetic field and energy particle distribution in this H\,{\sc ii} region.
\section{Summary and Conclusions}
\label{sec:conc}
In this paper, we present a detailed multi-wavelength analysis of an area ($\sim$0$\degr$.27 $\times$ 0$\degr$.27) around the H\,{\sc ii} region G18.88$-$0.49. 
The major outcomes of this observational work are the following:\\
$\bullet$ The H\,{\sc ii} region G18.88$-$0.49 is excited by an O-type main-sequence star (dynamical age $\sim$10$^{5}$ years for $n_\mathrm{0}$ = 10$^{3}$ cm$^{-3}$) and shows thermal emission.\\ 
$\bullet$ The H\,{\sc ii} region G18.937$-$0.434 exhibits a radio spectrum which we interpret as a composite spectrum of thermal and non-thermal emission. The thermal component is consistent with being powered by an early B-type main-sequence star.\\
$\bullet$ The observed non-thermal radio emission in G18.937$-$0.434 can be explained as synchrotron 
radiation generated by relativistic particles locally accelerated in shocked regions of gas interacting with an ambient 
magnetic field of $\sim 43~\mu$G.\\
$\bullet$ A shell-like feature (extension $\sim$12 pc $\times$ 7 pc) around the H\,{\sc ii} region G18.88$-$0.49 is investigated in the {\it Herschel} column density map. The distribution of molecular ($^{12}$CO, $^{13}$CO, C$^{18}$O, and NH$_\mathrm{3}$) gas at [60, 70] km s$^{-1}$ supports the presence of this shell-like feature.\\
$\bullet$ The shell-like feature (mass $\sim$2.9 $\times$10$^{4}$ M$_{\odot}$) contains high column density materials ($\ge$ 21 $\times$ 10$^{21}$ cm$^{-2}$) and at least four subregions (``A--D") having a mass range of $\sim$0.8--10.5 $\times$10$^{3}$ M$_{\odot}$.\\
$\bullet$ In the direction of the shell-like feature around the H\,{\sc ii} region G18.88$-$0.49, 8 Class~I YSOs \citep[mean age $\sim$0.44 Myr;][]{evans09}, 1 flat-spectrum source, and 11 ATLASGAL clumps (c3--c13) are traced. \\
$\bullet$ In six ATLASGAL clumps (i.e., c4, c5, and c8--c11), the thermal to non-thermal pressure ratio ranges from 0.04 to 0.12 with a mean of 0.09, and Mach number varies from 2.9 to 4.8 with an average of 3.5. 
These results indicate that these ATLASGAL clumps, distributed toward the {\it Herschel} subregions (``A--C"), are dominated by non-thermal pressure and supersonic non-thermal motions.\\ 
$\bullet$ One of the ATLASGAL clumps c11 ($L_{bol}$ $\sim$53.8 $\times$ 10$^{3}$ $L_\odot$; $M_\mathrm{clump}$ $\sim$4880 $M_\odot$) 
hosts the IRc of the 6.7 GHz methanol maser and a Class~I YSO, which is away from the H\,{\sc ii} region G18.88$-$0.49. 
The 6.7 GHz methanol maser is also associated with the EGO. The IRc appears to be a MYSO candidate \citep[stellar mass $\sim$7.7 M$_{\odot}$;][]{kerton13} associated with the molecular outflow.\\
$\bullet$ The early phases of star formation activities (including massive stars) are evident in the shell-like feature.\\
$\bullet$ Five parsec-scale filaments are found in the {\it Herschel} column density and molecular maps, 
and seem to be radially pointed to the dense parts of the shell-like feature. This configuration is considered as a ``hub-filament" system.\\
$\bullet$ Significant velocity gradients (0.8--1.8 km s$^{-1}$ pc$^{-1}$) are traced along the five filaments. 
These findings favor the molecular gas flow towards the central ``hub" along these filaments, feeding the central hub.\\

Overall, our findings are in agreement with a global non-isotropic collapse scenario as discussed in \citet{Motte+2018}. 
This scenario seems to successfully explain the observed star formation activities in our selected site around the H\,{\sc ii} region G18.88$-$0.49.
\acknowledgments  
We thank the anonymous reviewer for several useful comments and 
suggestions, which greatly improved the scientific contents of the paper. 
The research work at Physical Research Laboratory is funded by the Department of Space, Government of India. 
This work is based [in part] on observations made with the {\it Spitzer} Space Telescope, which is operated by the Jet Propulsion Laboratory, California Institute of Technology under a contract with NASA. 
This publication makes use of data from FUGIN, FOREST Unbiased Galactic plane Imaging survey with the Nobeyama 45-m telescope, a legacy project in the Nobeyama 45-m radio telescope. 
DKO acknowledges the support of the Department of Atomic Energy, Government of India, under project No. 12-R\&D-TFR-5.02-0200. SS  acknowledges the support of the Department of Science and Technology, Government of India, under project No. DST/INT/Thai/P-15/2019.
%
\begin{table*}
\setlength{\tabcolsep}{0.1in}
\centering
\caption{Physical parameters of 17 ATLASGAL dust clumps at 870 $\mu$m taken from \citet{urquhart18} (see Figure~\ref{fig2}h). 
All the clumps are located at a distance of 5.0 kpc. 
Table lists ID, Galactic coordinates ({\it l}, {\it b}), 870 $\mu$m integrated flux density ($S_\mathrm{870}$), radial velocity (V$_\mathrm{lsr}$), clump effective radius ($R_\mathrm{clump}$), dust temperature ($T_\mathrm{d}$), 
bolometric luminosity ($L_\mathrm{bol}$), clump mass ($M_\mathrm{clump}$), H$_\mathrm{2}$ column density ($N(\mathrm H_2)$), and 
average volume density ($n_{\mathrm H_2}$). Massive clumps ($>$ 10$^{3}$ M$_{\odot}$) are indicated by daggers.} 
\label{tab2}
\begin{tabular}{lcccccccccccr}
\hline 
 ID&   {\it l}     &  {\it b}      & $S_\mathrm{870}$  & V$_\mathrm{lsr}$   &$R_\mathrm{clump}$ &$T_\mathrm{d}$ &$L_\mathrm{bol}$ &$M_\mathrm{clump}$&$N(\mathrm H_2)$ & $n_{\mathrm H_2}$ \\  
      & (degree) & (degree)   &  (Jy)      & (km s$^{-1}$)     &     (pc)     & (K)&   (10$^{3}$ $\times$ $L_\odot$)& (10$^{2}$ $\times$ $M_\odot$) &(10$^{21}$ $\times$cm$^{-2}$)  &(10$^{3}$ $\times$ cm$^{-3}$)\\  
\hline
\hline 
      c1  &  18.763  &  -0.462  &   7.98  &   64.8  &   1.4  &   22.7  &    1.8   & 	9.1   &    7.9   &    1.1   \\
      c2  &  18.774  &  -0.459  &   4.22  &   67.5  &   0.9  &   21.2  &    0.9   & 	5.2   &    9.4   &    2.3   \\
      c3  &  18.798  &  -0.494  &   6.45  &   62.3  &   2.4  &   18.8  &    6.8   & 	9.6   &   16.2   &    0.2   \\
      c4  &  18.824  &  -0.467  &   2.65  &   63.0  &   0.6  &   22.7  &    4.1   & 	3.0   &   33.0   &    5.3   \\
      c5$\dagger$  &  18.823  &  -0.486  &  52.57  &   65.2  &   4.8  &   19.0  &   24.4   &   76.9   &   61.1   &    0.2   \\
      c6$\dagger$  &  18.843  &  -0.502  &  14.86  &   65.2  &   1.8  &   14.2  &    2.1   &   34.6   &   21.5   &    2.2   \\
      c7$\dagger$  &  18.854  &  -0.532  &  15.09  &   64.0  &   2.8  &   24.1  &   28.2   &   15.7   &   13.2   &    0.2   \\
      c8  &  18.863  &  -0.481  &   8.60  &   65.4  &   1.7  &   23.5  &    7.0   & 	9.2   &   23.9   &    0.7   \\
      c9$\dagger$  &  18.876  &  -0.489  &   8.20  &   65.5  &   1.5  &   19.3  &    2.4   &   11.7   &   39.6   &    1.2   \\
      c10$\dagger$  &  18.886  &  -0.509  &  17.78  &   66.2  &   2.9  &   25.0  &   23.9   &   17.6   &   18.1   &    0.2   \\
      c11$\dagger$  &  18.888  &  -0.474  &  43.05  &   65.8  &   4.7  &   22.7  &   53.8   &   48.8   &   86.5   &    0.2   \\
      c12  &  18.898  &  -0.499  &   1.99  &   66.2  &   0.6  &   18.3  &    1.0   & 	3.1   &   22.9   &    5.5   \\
      c13  &  18.898  &  -0.509  &   2.71  &   66.3  &   0.6  &   21.6  &    1.8   & 	3.3   &   18.2   &    5.8   \\
      c14  &  18.971  &  -0.404  &   2.91  &   66.9  &   0.6  &   13.7  &    0.1   & 	7.2   &   18.2   &   12.7   \\
      c15  &  18.948  &  -0.456  &   3.13  &   64.4  &   0.6  &   20.1  &    0.3   & 	4.2   &    8.7   &    7.5   \\
      c16$\dagger$  &  18.844  &  -0.376  &  10.51  &   61.0  &   2.5  &   14.9  &    1.3   &   22.6   &   30.1   &    0.5   \\
      c17$\dagger$  &  18.859  &  -0.416  &   5.94  &   67.8  &   1.8  &   15.3  &    0.4   &   12.2   &   18.6   &    0.8   \\
\hline	     
\end{tabular}
\end{table*}
\begin{table*}
\setlength{\tabcolsep}{0.1in}
\centering
\caption{Physical parameters of six ATLASGAL clumps derived using the NH$_\mathrm{3}$ line data \citep[from][]{wienen12}. 
Table contains the ATLASGAL clump ID, NH$_\mathrm{3}$ (1,1) line velocity (V$_\mathrm{lsr}$), NH$_\mathrm{3}$ (1,1) line 
width ($\Delta V$), kinematical temperature ($T_\mathrm{kin}$), sound speed ($a_\mathrm{s}$), thermal velocity dispersion ($\sigma_\mathrm{T}$), non-thermal velocity dispersion ($\sigma_\mathrm{NT}$), Mach number ($M$ = $\sigma_\mathrm{NT}$/$a_\mathrm{s}$), and 
ratio of thermal to non-thermal gas pressure ($R_\mathrm{p}$ = ${a^2_\mathrm{s}}/{\sigma^2_\mathrm{NT}}$) (see Table~\ref{tab2} and also filled hexagons in Figure~\ref{fig2}h).}
\label{tab3}
\begin{tabular}{lccccccccccccc}
\hline 										    	        			      
  clump ID          &   V$_\mathrm{lsr}$      &	$\Delta V$ &      $T_\mathrm{kin}$          & $a_\mathrm{s}$        & $\sigma_\mathrm{T}$ &$\sigma_\mathrm{NT}$  & $M$ & $R_\mathrm{p}$  \\ 
             &     (km s$^{-1}$) &   (km s$^{-1}$)    &     (K)         &   (km s$^{-1}$)    &   (km s$^{-1}$) &   (km s$^{-1}$)& ($\sigma_\mathrm{NT}$/$a_\mathrm{s}$) &  (${a^2_\mathrm{s}}/{\sigma^2_\mathrm{NT}}$) \\ 
\hline 
   c4   &     63.0    &    1.80  &     19.91     &	 0.26	&      0.10  &     0.76  &     2.91   &     0.12 \\
   c5   &     65.2    &    2.06  &     20.81     &	 0.27	&      0.10  &     0.87  &     3.26   &     0.09 \\
   c8   &     65.4    &    2.87  &     18.83     &	 0.25	&      0.09  &     1.22  &     4.79   &     0.04 \\
   c9   &     65.5    &    1.90  &     22.39     &	 0.28	&      0.10  &     0.80  &     2.89   &     0.12 \\
   c10  &     66.2    &    2.28  &     23.57     &	 0.28	&      0.11  &     0.96  &     3.39   &     0.09 \\
   c11  &     65.8    &    2.85  &     27.99     &	 0.31	&      0.12  &     1.20  &     3.90   &     0.07 \\
\hline          		
\end{tabular}			
\end{table*}			

\end{document}